\documentclass[a4paper,fleqn,usenatbib]{mnras}
\usepackage[T1]{fontenc}
\usepackage{ae,aecompl}

\usepackage{graphicx}	
\usepackage{amsmath}	
\usepackage{amssymb}	
\usepackage{makecell}
\usepackage{multirow}
\usepackage{hyperref}

\newcommand{\kms}{km s$^{-1}$}
\newcommand{\dego}{$^\circ$}

\title[The nascent wind of AGB star R Doradus]{The nascent wind of AGB star R Doradus: evidence for a recent episode of enhanced mass loss}
\author[P.T. Nhung et al.]{{P.T. Nhung\thanks{E-mail: pttnhung@vnsc.org.vn}, D.T. Hoai, P. Tuan-Anh, P. Darriulat, P.N. Diep,}
\newauthor{ N.T. Phuong, T.T. Thai} 
\\
Department of Astrophysics, Vietnam National Space Center (VNSC), Vietnam Academy of Science and Technology (VAST),\\
18 Hoang Quoc Viet, Cau Giay, Ha Noi, Vietnam\\
}
\date{Accepted XXX. Received YYY; in original form ZZZ}

\pubyear{2019}
\begin{document}
\label{firstpage}
\pagerange{\pageref{firstpage}--\pageref{lastpage}}
\maketitle

\begin{abstract}
 We analyse ALMA observations of the SO($J_K=6_5-5_4$) emission of the circumstellar envelope of oxygen-rich AGB star R Dor, probing distances between 20 and 100 au from the star where the nascent wind is building up. We give evidence for the slow wind to host, in addition to a previously observed rotating disc, a radial outflow covering very large solid angles and displaying strong inhomogeneity both in direction and radially: the former takes the form of multiple cores and the latter displays a radial dependence suggesting an episode of enhanced mass loss having occurred a century or so ago.

\end{abstract}



\section{Introduction}  \label{sec1}

In a recent study of the circumstellar envelope of oxygen-rich AGB star R Dor \citep{Hoai2019}, we commented on the detection of high Doppler velocity components confined near the line of sight and reaching up to nearly $\sim$20 \kms, critically comparing interpretations in terms of narrow gas streams with interpretations in terms of artefacts of improper continuum subtraction.  In the present paper, we study instead the slower wind, exclusive of higher Doppler velocities. Earlier studies by \citet{Khouri2016}, \citet{Danilovich2016}, \citet{DeBeck2018}, \citet{VandeSande2018}, \citet{Decin2018}, \citet{Homan2018} and \citet{Vlemmings2018} have contributed a considerable amount of detailed information of relevance to the physico-chemistry and dynamics of both dust and gas in this slower wind, however without attempting a detailed description of the morpho-kinematics, which is the aim of the present paper.

R Dor, an oxygen-rich AGB star, is a semi-regular variable of the SRb type, belonging to spectral class M8IIIe, having an initial mass between 1 and 2 M$_\odot$, a mass loss rate of $\sim$1.6 10$^{-7}$ M$_\odot$yr$^{-1}$ and a temperature of $\sim$3058 K \citep{Dumm1998}; it is close to the Sun, at a distance of only $\sim$59 pc \citep{Knapp2003}; it displays no technetium in its spectrum \citep{Lebzelter1999} and has a $^{12}$CO/$^{13}$CO abundance ratio of $\sim$10 \citep{Ramstedt2014}. It has a dual period of 175 and 332 days \citep{Bedding1998} and its infrared emission above black body between 1 and 40 $\mu$m wavelength shows an enhancement of aluminium oxide and melilite in its dust \citep{Heras2005}.

Very little is known of R Dor at distances in excess of $\sim$150 au from the star, except that its wind has an approximate terminal velocity of 5 to 6 \kms\ \citep{VandeSande2018}, a result obtained from modelling a number of single dish observations assuming spherical symmetry. On the contrary, at short distances from the star, it has been observed by ALMA with a spatial resolution of $\sim$2.2 au, resolving a stellar disc of $\sim$3.6 au in diameter and giving evidence for rotation with a velocity of (1.0$\pm$0.1)/$\sin i$ \kms\ at the stellar surface about an axis projecting 7\dego$\pm$6\dego\ east of north on the sky plane \citep{Vlemmings2018}; however, the angle $i$ between the rotation axis and the line of sight is unknown. The dust has been observed at the VLT with a resolution of 1.2 au \citep{Khouri2016} and the gas envelope has been probed by ALMA with a resolution of 9 au up to some 20 au from the star \citep{Decin2018, Homan2018}, the presence of a south-eastern ``blue-blob'', $\sim$15 au from the star, being interpreted as suggesting the existence of a companion; strong absorption of the stellar disc emission has been observed in the blue-shifted hemisphere; evidence has been obtained for a rotating gas disc surrounding the star, with an outer radius of $\sim$25 au and an axis making an angle of $\sim$20\dego$\pm$20\dego\ with the plane of the sky and projecting on it 25\dego$\pm$5\dego\ east of north \citep{Homan2018}.

Sulphur monoxide in the R Dor environment has been studied in detail by \citet{Danilovich2016} with the result that sulphur monoxide and dioxide share the quasi-totality of sulphur-bearing species and display a Gaussian radial profile centred on the star and extending up to $\sim$80 au (HWHM) from the star.

In what follows, we explore the morpho-kinematics of the circumstellar envelope using ALMA observations of the SO($J_K=6_5-5_4$) emission, taking advantage of the large time on source, more than six times longer than for line emissions studied earlier. In particular we aim at obtaining a better description of the roles played by rotation and expansion and to place additional constraints on the mechanisms governing the building-up of the nascent wind, including radiation pressure on dust grains, binarity and pulsations.

\section{Observations and data reduction} \label{sec2}

\begin{figure}
  \includegraphics[width=0.48\textwidth,trim=.5cm 1.cm 0.5 0,clip]{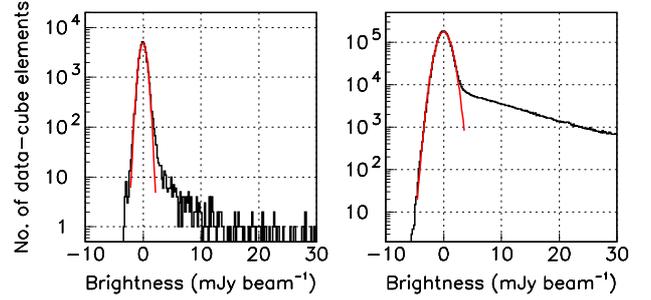}
  \caption{Distributions of the measured brightness for continuum (left) and SO line (right) emissions. Gaussian fits (red curves) give noise levels (standard deviation) of 0.6 mJy/beam and 1.1 mJy/beam respectively. The continuum distribution is for $R<3$ arcsec and the line distribution for $|x|<2$ arcsec, $|y|<2$ arcsec and $|V_z|<25$ \kms.}
  \label{fig1}
\end{figure}

The data, retrieved from ALMA archives and reduced by the ALMA staff, are from project 2017.1.00824.S observed in December 2017 for $\sim$2.7 hours on source in band 6 with an average of 45 antennas. The data-cube covers $\sim$4$\times$4 arcsec$^2$ and 40 \kms\ with elements of 28$\times$28 mas$^2$ and 0.29 \kms. In addition to continuum emission, we detect the transition between two fine structure rotational states of the SO molecule in its vibrational ground-state: $\nu=0$ , $J_K=6_5$ to $5_4$ with frequency of 251.826 GHz. The beam is circular with a FWHM of 0.15 arcsec (9 au), the excellent spatial resolution implying a steep radial acceptance. The continuum contribution has been subtracted in the $uv$ plane. Figure \ref{fig1} displays the distribution of the brightness $f$ measured in each data-cube element, giving evidence for a Gaussian noise of 1.1 mJy/beam (1$\sigma$). Also shown is the distribution of the brightness measured from continuum emission, with a Gaussian noise of 0.6 mJy/beam (1$\sigma$).

We use coordinates having the $y$ axis pointing north and the $x$ axis pointing east; the $z$ axis points away from us, parallel to the line of sight, and the origin of coordinates is taken at the centre of continuum emission (measured 23 mas east and 34 mas north from the position used by the ALMA staff for data reduction). We call $R$ the projection on the plane of the sky of the distance to the star, $R=\sqrt{x^2+y^2}$. Doppler velocities ($V_z$) are referred to a local standard of rest velocity of 7.0 \kms. Position angles $\omega$ are measured on the sky plane counter-clockwise from north.

\section{Continuum and line emissions: an overview} \label{sec3}  

\begin{figure*}
  \includegraphics[width=0.95\textwidth,trim=0.cm 1.cm 0 0,clip]{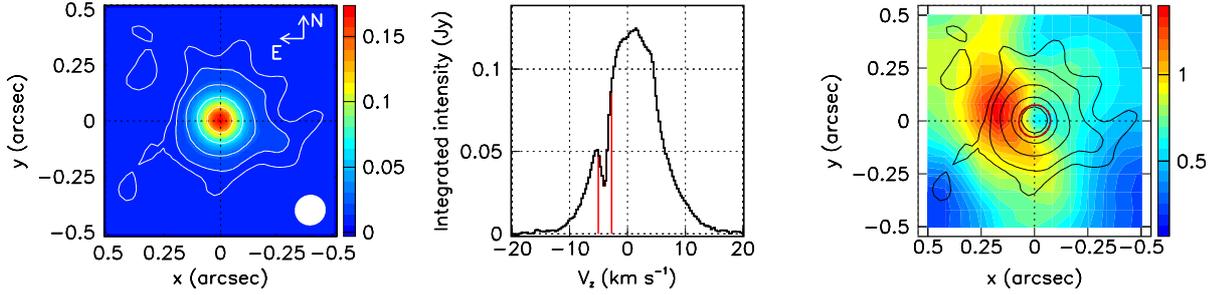}
  \caption{Left: map of continuum emission. A standard deviation of 80 mas (FWHM of 190 mas compared with 150 mas for the beam) is measured in each of $x$ and $y$. The colour scale is in units of Jy beam$^{-1}$. The beam is shown in the lower right corner of the panel. Middle: Doppler velocity spectrum of the line emission over the stellar disc ($R<90$ mas) shown as a red circle in the right panel. Right: map of the line intensity integrated over Doppler velocities between $-5$ and $-3$ \kms, shown with red lines in the middle panel. Also shown are contours of the continuum map. The colour scale is in units of Jy beam$^{-1}$ \kms.}
  \label{fig2}
\end{figure*}

The map of continuum emission is displayed in the left panel of Figure \ref{fig2}. The stellar disc, measured by \citet{Vlemmings2018} with a radius of 1.8 au (30 mas) is not resolved but absorption of its emission by the blue-shifted wind is visible at Doppler velocities between $-5$ and $-3$ \kms, corresponding to the dominant radial velocity of the slow wind in the blue-shifted hemisphere. This is illustrated in the middle and right panels of Figure \ref{fig2} as depressions of the Doppler velocity spectrum integrated over the star location or of the intensity map integrated over the $-5$ to $-3$ \kms\ interval. The importance of continuum emission from the star at projected distances $R<0.2$ arcsec complicates the study of the line emission in this region.

\begin{figure*}
  \includegraphics[width=0.95\textwidth,trim=0.cm 0.5cm 0 0,clip]{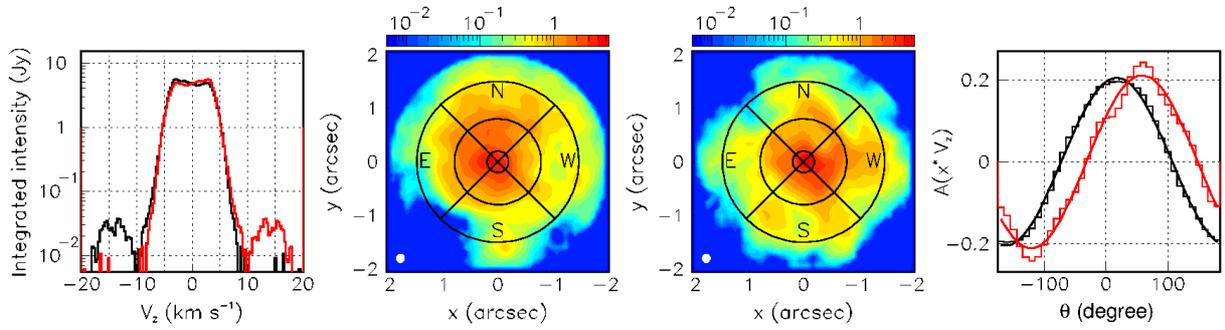}
  \caption{Left: Doppler velocity spectrum integrated over $0.2<R<1.5$ arcsec (black) and its mirror-symmetric with respect to the origin (red). Middle: intensity maps integrated over $-25< V_z <0$ \kms\  (middle left) and $0<V_z<25$ \kms\ (middle right). The colour scales are in units of Jy beam$^{-1}$ \kms. The beams are shown in the lower left corners of the panels. Right: Dependence on $\theta$ of the asymmetry $A(x^*V_z)$ for the inner (black, $0.2<R<0.8$ arcsec) and outer (red, $0.8<R<1.5$ arcsec) radial regions. The curves are fits of the form $A_0\cos(\theta-\theta_0)$ with $(A_0, \theta_0)$=(0.21, 17\dego) in the inner radial region and $(A_0,\theta_0)$=(0.21, 58\dego) in the outer radial region.}
  \label{fig3}
\end{figure*}

The line emission displays a nearly symmetric profile (Figure \ref{fig3} left) with a flat top of $\pm$3.5 \kms\ bracketed on each side by a decrease to zero covering $\sim$2.5 \kms.

Intensity maps are displayed in the middle-left and middle-right panels of Figure \ref{fig3} for the blue and red-shifted hemispheres separately. They cover the range where the SO abundance relative to H$_2$ is known to be significant \citep{Danilovich2016}. They reveal important anisotropy of the data-cube, which we quantify by considering separately the radial regions $0.2<R<0.8$ arcsec and $0.8<R<1.5$ arcsec in the Doppler velocity interval $|V_z|<6$ \kms. In each of these regions we split the data-cube in eight octants depending on the sign of $V_z$ and on the position angle (sectors N, E, S and W shown in Figure \ref{fig3}). We define $f_n$ as the intensity integrated in octant $n$ and $F_n$ as its value normalised to the mean $<f>$:\\
$F_n=f_n/<f>$ with $<f>=\sum\limits_{n=1}^8 f_n/8$; 	$ \sum\limits_{n=1}^8 F_n=8$ \\
We measure anisotropies with an amplitude $k$ and a normalised distribution $g_n$, defined as follows:\\
$F_n=1+k g_n$ with $<$$|g_n|$$>$=$\sum\limits_{n=1}^8 |g_n|/8$=1; $\sum\limits_{n=1}^8 |g_n|$=8; 	$\sum\limits_{n=1}^8 g_n$=0.

Their values, listed in Table \ref{tab1}, reveal important differences between the inner and outer regions, although the global anisotropies, $k$=20\% and 23\%, respectively, are similar. The inner region, $0.2<R<0.8$ arcsec, displays strong excess of the blue-shifted E sector and red-shifted W sector over the red-shifted E sector and blue-shifted W sector, as expected from rotation about an axis projecting between 0\dego\ and 30\dego\ east of north, as observed by \citet{Homan2018} and, on the star, by \citet{Vlemmings2018}. This is best seen from the angular dependence of the asymmetry $A(x^*V_z)$ between positive and negative $x^*V_z$, $A(x^*V_z)= [\sum{f(x^*V_z>0)}-\sum{f(x^*V_z<0)}]/[\sum{f(x^*V_z>0)}+\sum{f(x^*V_z<0)}]$, where $x^*=x\cos\theta-y\sin\theta$, meaning that $x^*$ points $\theta$ degrees counter-clockwise from east. In the case of rotation, $A(x^*V_z)$ is maximal when $x^*$ points to a direction perpendicular to the projection of the rotation axis on the sky plane. As illustrated in the right panel of Figure \ref{fig3} for each radial region separately, the dependence of $A(x^*V_z)$ over $\theta$ is well described in the inner radial region by a sine wave of the form $A(x^*V_z)=0.21\cos(\theta-17$\dego), corresponding to rotation about an axis projecting on the plane of the sky 17\dego\ east of  north, compared with 25\dego$\pm$5\dego\ obtained by \citet{Homan2018} below 25 au ($\sim$0.4 arcsec) and 7\dego$\pm$6\dego\ obtained by \citet{Vlemmings2018} on the star.

    In the outer region, the situation is more complex than in the inner region. The asymmetry observed in the latter between S and N sectors is now strongly amplified in the blue-shifted hemisphere where the N sector exceeds the S sector by as much as 150\%. The sine wave fit to $A(x^*V_z)$ is not as good as in the inner region and gives a very large phase-shift of 58\dego, possibly revealing significant anisotropic radial expansion.
    
The main sources of anisotropy are rotation, expansion and local excesses or depletions that may be caused, for example, by turbulence, shocks or the presence of a stellar or planetary companion. Rotation about an axis projecting at position angle $\theta$ on the plane of the sky \citep{Diep2016, Nhung2018} produces a sine wave modulation of the $xV_z$ asymmetry that cancels for $\theta=\pm$90\dego and has amplitude proportional to $\sin i$, where $i$ is the angle of the rotation axis with the line of sight (the $z$ axis). Radial expansion does not produce any anisotropy if it is spherical but does if it is axi-symmetric about the axis of rotation and enhanced near the poles or near the equator. When enhanced near the poles, it also produces a sine wave modulation of the $xV_z$ asymmetry, that is instead maximal for $\theta=\pm$90\dego. The combination of bipolar expansion and rotation about a same axis produces a sine wave with a phase-shift $\theta_0$ that depends on the relative importance of expansion with respect to rotation \citep{Diep2016, Nhung2018}. 

\begin{table*}
  \caption{Anisotropy of the SO data-cube.}
  \label{tab1}
  \begin{tabular}{ccccccccccc}
    \hline
    $<f>$&\multirow{2}{*}{$k$}&\multirow{2}{*}{}&\multicolumn{4}{c}{$V_z<0$}&\multicolumn{4}{c}{$V_z>0$}\\
    Jy\,\kms& &&N&E&S&W&N&E&S&W\\
    \hline
    \multicolumn{11}{c}{$0.2<R<0.8$ arcsec}\\
      \hline
      \multirow{3}{*}{2.69}&\multirow{3}{*}{20\%}&$f_n$&3.18&3.64&2.56&1.93&2.17&2.06&2.52&3.44\\
     
      &&$F_n$&1.18&1.35&0.95&0.72&0.81&0.77&0.94&1.28\\
     
      &&$g_n$&0.89&1.73&$-$0.25&$-$1.38&$-$0.94&$-$1.14&$-$0.30&1.38\\
      \hline
      \multicolumn{11}{c}{$0.8<R<1.5$ arcsec}\\
      \hline
      \multirow{3}{*}{2.60}&\multirow{3}{*}{23\%}&$f_n$&4.15&2.62&1.66&2.47&2.30&1.65&2.74&3.22\\
      
      &&$F_n$&1.60&1.01&0.64&0.95&0.88&0.63&1.05&1.24\\
      
      &&$g_n$&2.67&0.04&$-$1.60&$-$0.22&$-$0.53&$-$1.64&0.22&1.07\\
      \hline
  \end{tabular}
\end{table*}

\begin{figure}
  \includegraphics[width=0.49\textwidth,trim=0.5cm 0.cm 0 0.5,clip]{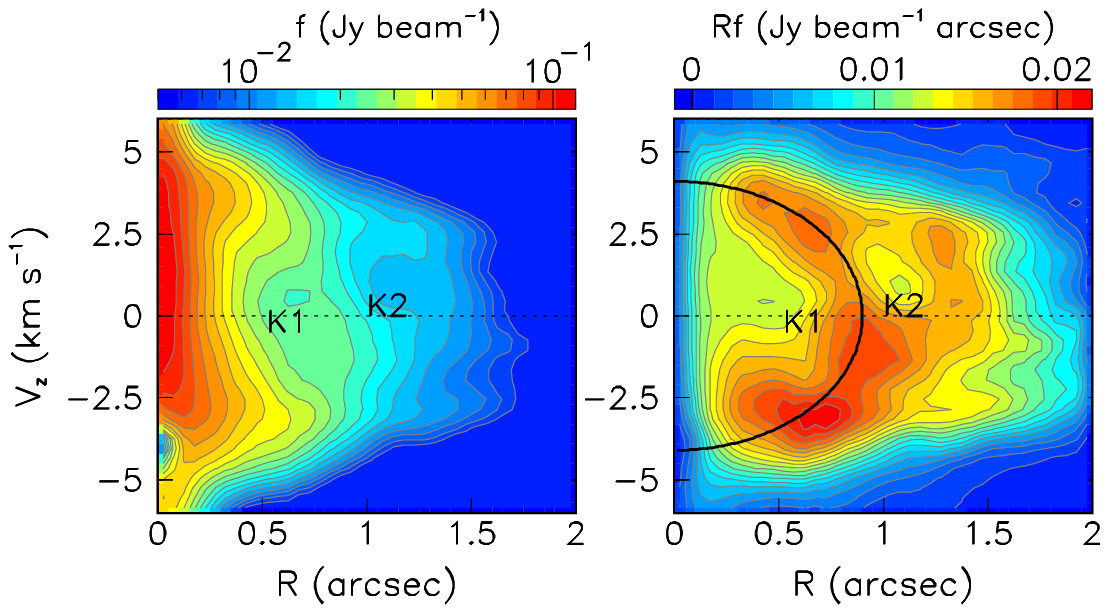}
  \caption{P-V maps of the brightness $f$ (left panel) and of its product $Rf$ by the projected distance $R$ (right panel) in the $V_z$ vs $R$ plane, averaged over position angle $\omega$.}
  \label{fig4}
\end{figure}

The presence of depletions that cannot be assigned to rotation or radial expansion in the R Dor data-cube was noted by \citet{Hoai2019} for Doppler velocities between $-$1 and $+$3 \kms; they can be expected to contribute significant anisotropy that will complicate the description of the data-cube in terms of rotation and expansion. Additional evidence for significant depletions of the data-cube is presented in Figure \ref{fig4}, which displays P-V maps in the $V_z$ vs $R$ plane, averaged over position angle $\omega$. They map the brightness $f$  and its product by the projected distance $R$, the latter giving a better representation of the situation at large distances (an isotropic emissivity inversely proportional to the square of the distance to the star and multiplied by $R$ would produce a uniform intensity). It shows two rings of low emission at mean projected distances of $\sim$0.6 and $\sim$1.1 arcsec from the star having mean Doppler velocities of $\sim$0.5 and $\sim$1 \kms, labelled K1 and K2 respectively. Such morphology is likely to reveal episodic enhancements of the mass loss, the history of which is recorded in the data-cube. Figure \ref{fig4} suggests such an enhancement on an approximately ellipsoidal shell having its axis on the $V_z$ axis, where it reaches between $\sim\pm$4 to $\pm$5 \kms, and a radius a little below 1 arcsec in the $V_z=0$ plane.

The overview of the structure of the data-cube presented in this section has demonstrated the complex structure of the morpho-kinematics at stake, and revealed the difficulty of producing a reliable description of the physico-chemical mechanisms that cause it. Before attempting such description it is therefore essential to obtain as detailed as possible a picture of the structure of the data-cube.

\section{A 3-D description of the data-cube} \label{sec4}

\begin{figure*}
 \includegraphics[height=0.35\textheight,trim=0.3cm 0.5cm .7cm 0.5cm,clip]{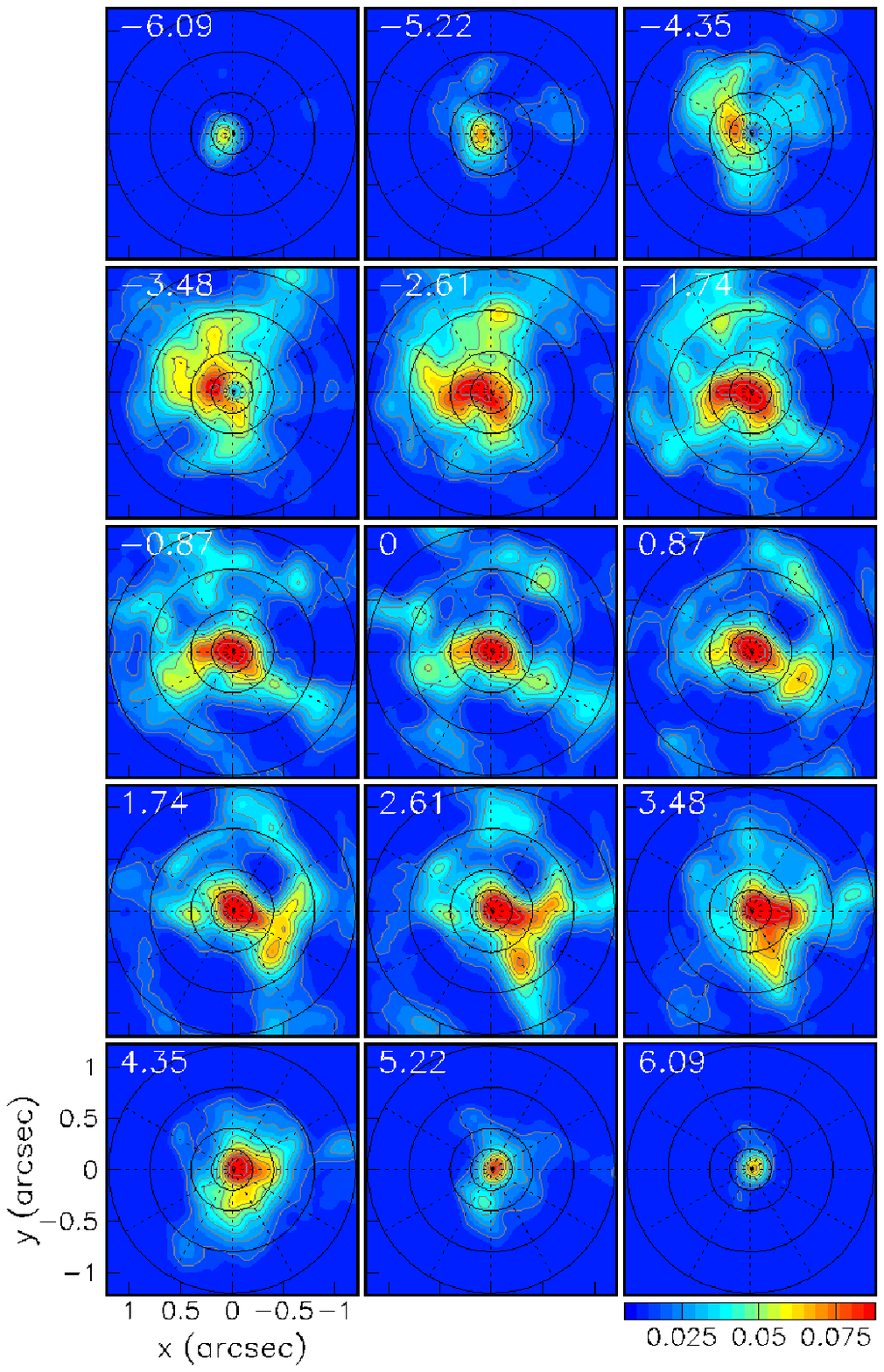}
 \includegraphics[height=0.367\textheight,trim=0.3cm 0.6cm 1.5cm 1cm,clip]{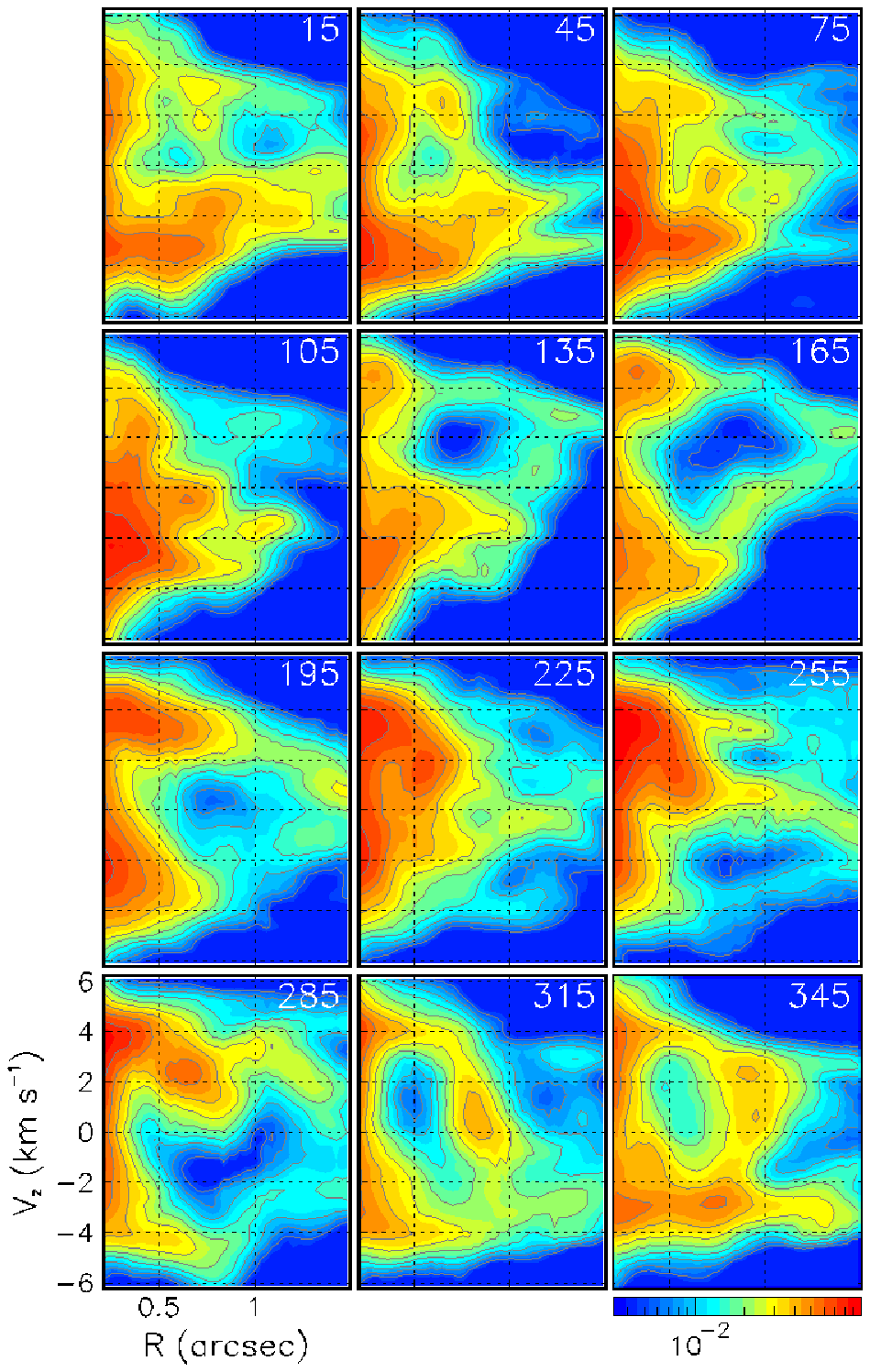}
 \includegraphics[height=0.35\textheight,trim=0.cm 0.5cm 0.cm 0.5cm,clip]{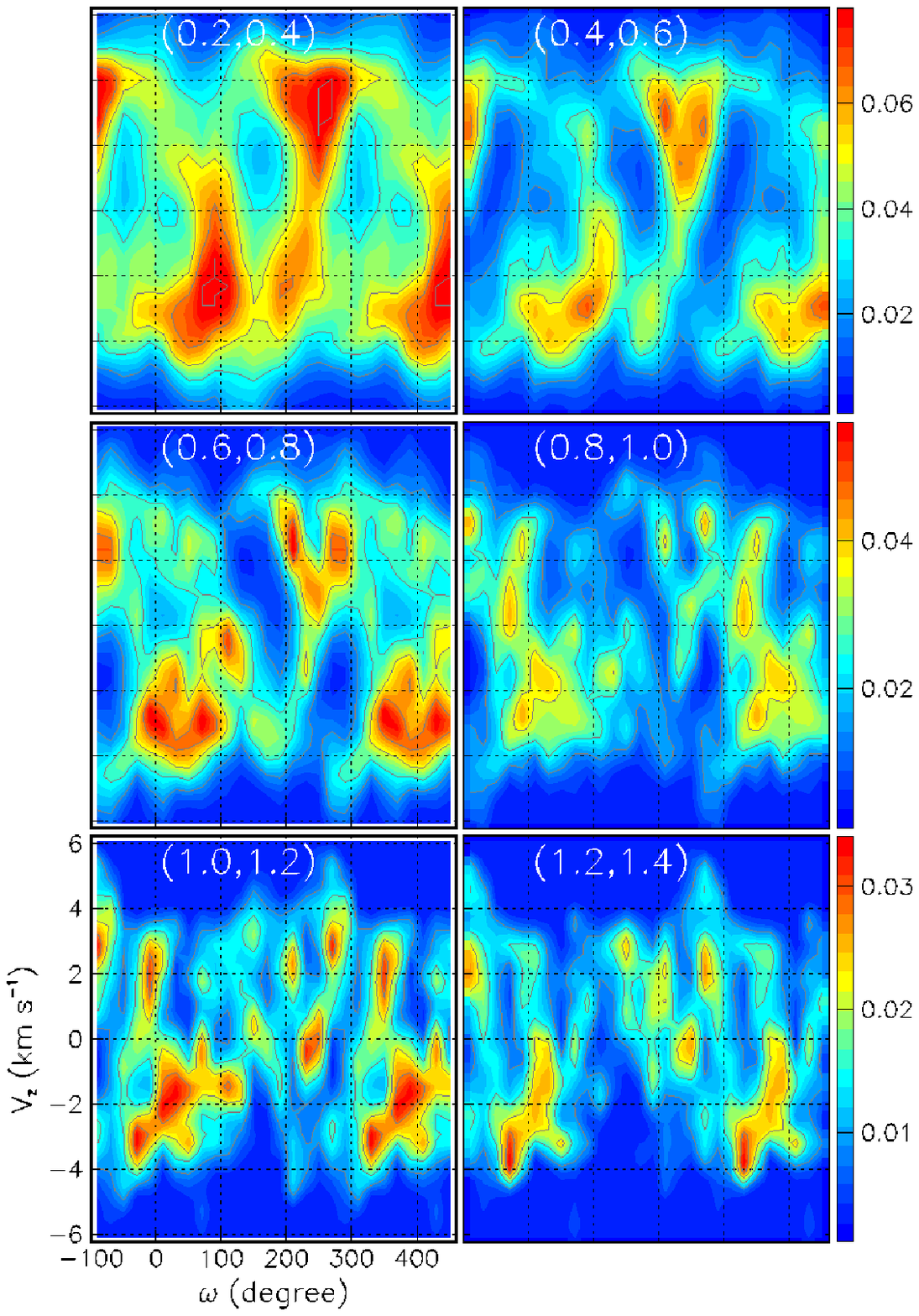}
  \caption{Projections of the data-cube on three different planes: as channel maps in $y$ vs $x$ in bins of $V_z$ (left); as P-V maps in $V_z$ vs $R$ in bins of $\omega$ (middle); and as P-V maps in $V_z$ vs $\omega$ in bins of $R$ (right). Angles are in degrees, distances in arcsec and velocities in \kms. The colour scales are in units of Jy beam$^{-1}$.}
  \label{fig5}
\end{figure*}

Figure \ref{fig5} displays projections of the data-cube on three different planes: as channel maps in $y$ vs $x$ in bins of $V_z$ (left); as P-V maps in $V_z$ vs $R$ in bins of $\omega$ (middle); and as P-V maps in $V_z$ vs $\omega$ in bins of $R$ (right). While the following analysis uses much finer segmentations of the data-cube presented as \href{https://vnsc.org.vn/dap/files/supp19.01.pdf}{supplementary material} to the article, Figure \ref{fig5} is sufficient to illustrate the main points. Immediately apparent from inspection of the figure is the complexity of the morpho-kinematics at stake and the difficulty that it implies in attempting a reliable de-projection.  

The latter of the three projections, the $V_z$ vs $\omega$ P-V maps, invites some simple remarks. It is dominated up to $R\sim0.8$ arcsec by an enhancement centred near $\omega=$270\dego and $V_z=+2.5$ \kms\ and an enhancement centred near $\omega=$90\dego and $V_z=-2.5$ \kms, suggesting two obvious possible scenarios: rotation about an axis projecting on the sky plane near the $y$ axis or radial expansion directed along an axis projecting on the sky plane near the $x$ axis. Section \ref{sec4.1} below studies these enhancements in some detail and Section \ref{sec4.2} studies the central region that hosts the rotating disc described by \citet{Homan2018}. Also apparent on the $V_z$ vs $\omega$ P-V maps is the presence of important islands of low emissions, suggesting the presence of cavities in the data-cube as has been shown in Figure \ref{fig4}. Section \ref{sec4.3} studies such a cavity that circles the star at small Doppler velocities, $|V_z|<\sim2$ \kms.  

\subsection{Outflows: rotation and expansion}\label{sec4.1}

\begin{figure}
  \includegraphics[width=0.48\textwidth,trim=0.5cm .5cm 0 1.,clip]{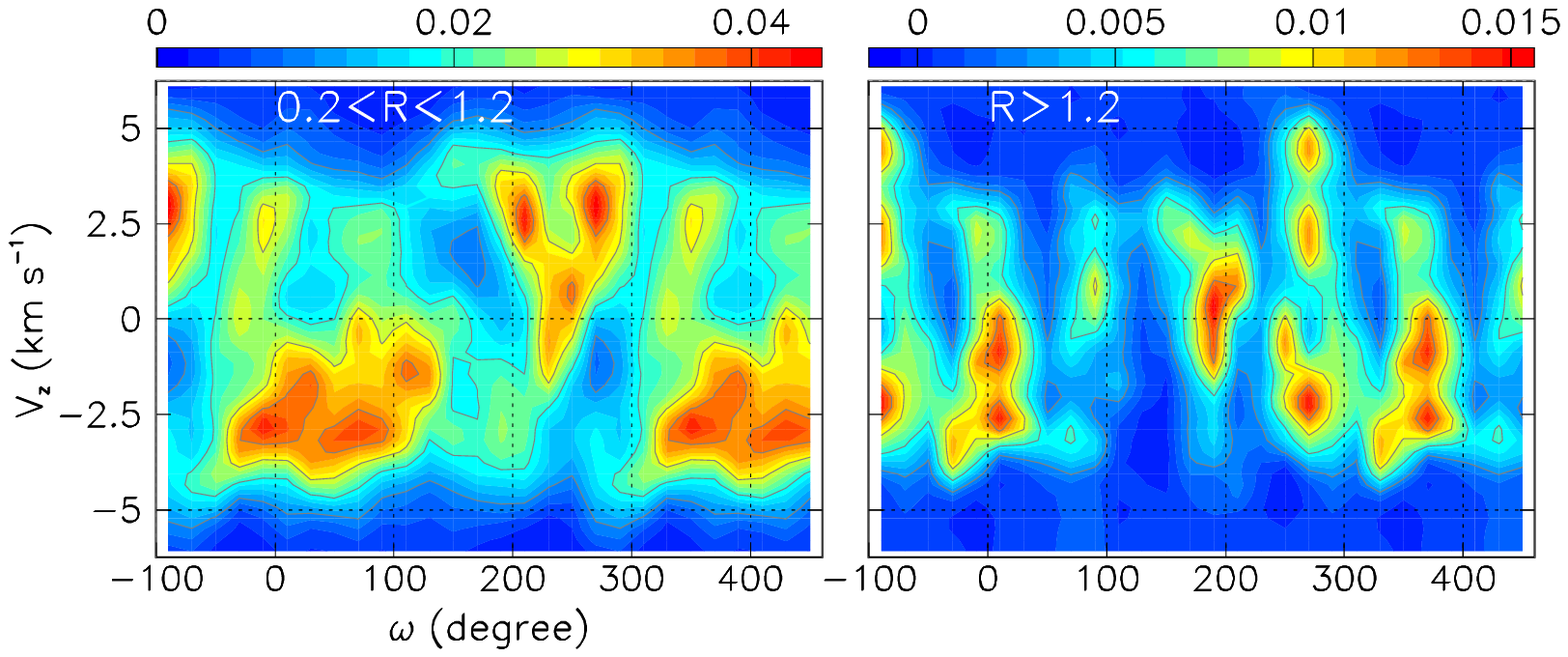}
  \caption{P-V maps in the $V_z$ vs $\omega$ plane for $0.2<R<1.2$ arcsec (left) and $R>1.2$ arcsec (right). The colour scales are in units of Jy beam$^{-1}$.}
  \label{fig6}
\end{figure}

Figure \ref{fig6} displays $V_z$ vs $\omega$ P-V maps, averaged over $0.2<R<1.2$ arcsec and $R>1.2$ arcsec, respectively. The former is dominated by two enhancements, one blue-shifted, in the eastern hemisphere and the other red-shifted, in the western hemisphere. The latter shows, at larger distances from the star, a more complex pattern.

We limit the discussion of the present section to the study of the main enhancements of emission observed within 1.2 arcsec projected distance from the star. A number of features plead against interpretations in terms of pure rotation as well as in terms of pure expansion.

Both enhancements display inhomogeneity in the form of three clear cores for the red-shifted enhancement and possibly four cores for the blue-shifted enhancement. The three-core structure of the former is invariant on the $V_z$ vs $\omega$ P-V map up to $R\sim1$ arcsec, beyond which distance the $V_z$ values of each of the three cores decrease by approximately 1 \kms\ over the following 0.5 arcsec. The approximate invariance of the three-core pattern as a function of $R$ favours an interpretation in terms of expansion, characterized by projective geometry, over an interpretation in terms of rotation: on average, the core values of $\omega$ and $V_z$ do not vary by more than $\pm$5\dego\ and 0.4 \kms\ respectively. In the case of rotation, one expects inhomogeneity to display circular or spiral patterns rather than radial. On the contrary, in a radial outflow, one expects the same features to reproduce at different distances from the star. For both rotation and expansion, the decrease of $V_z$ observed beyond $R\sim1$ arcsec would mean a slowing down of the respective velocities; however, such a slowing down cannot cause a change of sign of $V_z$ as observed for the middle core of the red-shifted enhancement, from $\sim1$ \kms\ to about $-0.5$ \kms, excluding interpretations in terms of pure rotation as well as in terms of pure radial expansion.

More generally, whether assuming rotation or expansion, the presence of strong inhomogeneity is not amenable to simple interpretation. In particular, in the present case, the enhancements cover very large angles, in excess of 90\dego, and Doppler velocity intervals of some 4 \kms; the cores of the red-shifted enhancement are well separated and some 30\dego\ and 2 \kms\ apart from each other.

\begin{figure*}
  \includegraphics[height=0.33\textheight,trim=1.cm .5cm 0.5cm 0.8cm,clip=true]{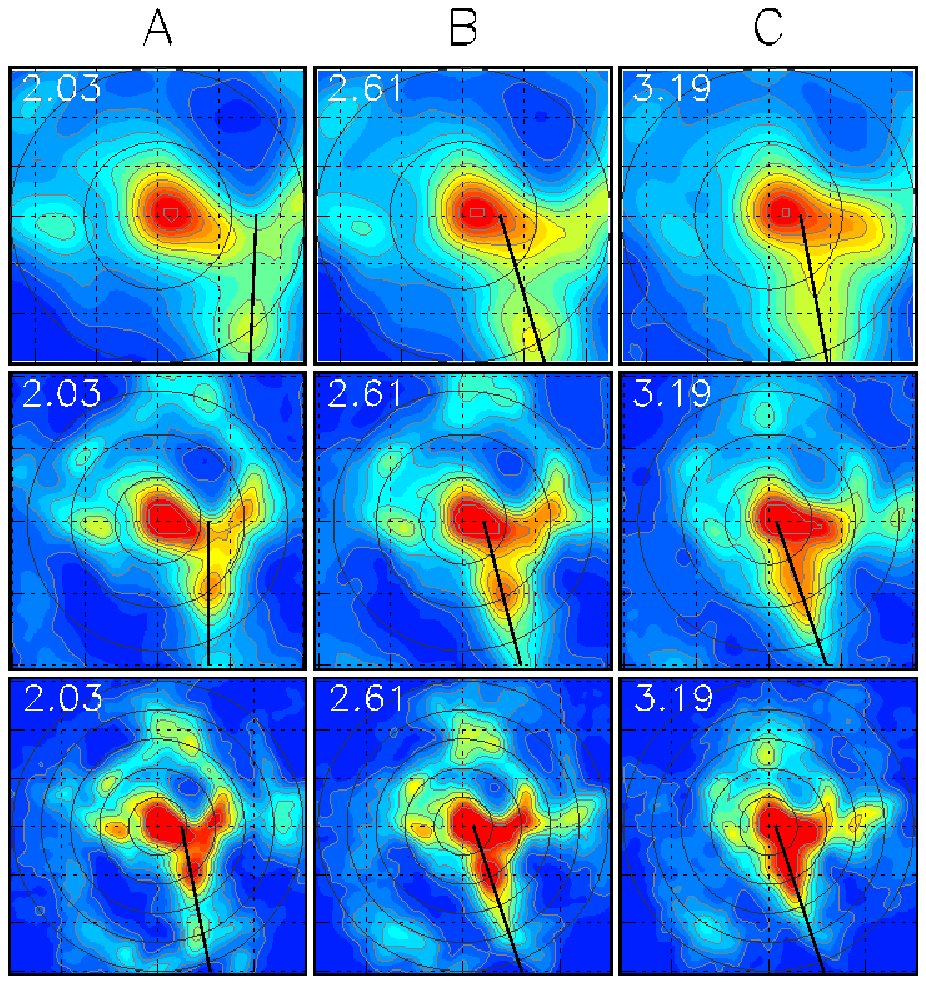}
  \includegraphics[height=0.33\textheight,trim=0.5cm .5cm 0.5cm 0.8cm,clip=true]{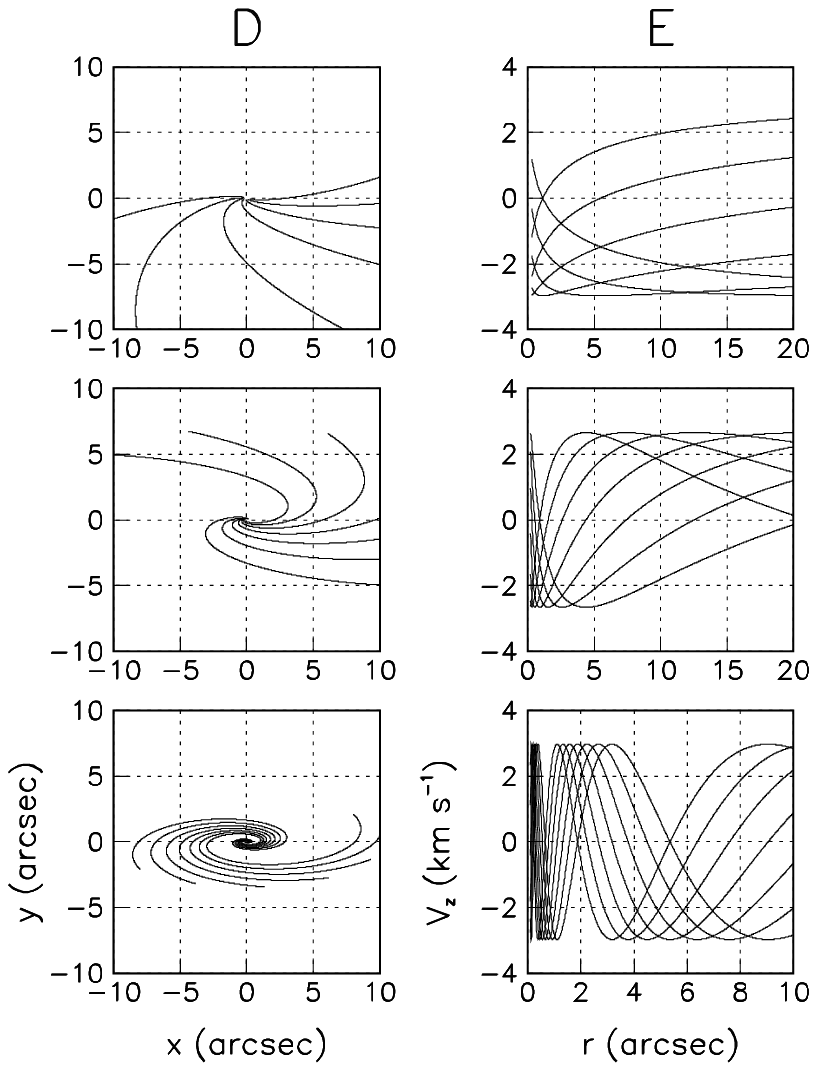} 
  \caption{Left panels: Birth of the southern core of the red-shifted enhancement mapped at three different Doppler velocities (2.03, 2.61 and 3.19 \kms, left, centre and right, respectively). The upper, middle and lower rows covers $\pm$0.6 arcsec, $\pm$1 arcsec and $\pm$1.5 arcsec in $x$ and $y$, respectively, with circles at 0.3, 0.6, 0.9, 1.2 and 1.5 arcsec in radius. The lines are drawn by hand as attempts to estimate extrapolations of the average emission to the $x$ axis. Right panels: trajectories (left column) and $V_z$ vs $R$ P-V diagrams (right column) calculated for $i$=70\dego\ and ($V_{rot}$, $V_{exp}$)=(1,3), (2,2) and (3,1) \kms, respectively.}
  \label{fig7}
\end{figure*}

Another remarkable feature is the failure of the observed enhancements to point to the star when observed at short distances to it, contrary to what should be expected in the case of pure radial expansion. This is illustrated, as an example, in Figure \ref{fig7}, which maps the birth of the southern core of the red-shifted enhancement at three different Doppler velocities (2.03, 2.61 and 3.19 \kms) and three different radial scales ($\pm$0.6, $\pm$1 and $\pm$1.5 arcsec). Extrapolation to the $x$ axis points some 0.2 arcsec west of the star on average, the more so when extrapolated from shorter distances and also the more so at lower Doppler velocities. This suggests an interpretation in terms of combined rotation and expansion. The trajectory of a gas volume having radial expansion velocity $V_{exp}$ and rotation velocity $V_{rot}$ is locally a logarithmic spiral that projects on the plane of the sky as a curved line. If gravitational pull from the star were the only force acting on it, both rotation and expansion velocity would decrease in inverse proportion to the square root of the distance. For expansion to dominate at large distances, as suggested from Figure \ref{fig6} and required in order to produce outflows on a large scale, the expansion velocity needs to significantly exceed the rotation velocity at larger distances, probably because the rotation velocity decreases faster than the expansion velocity as a function of distance.

In order to illustrate the complexity of the kinematics resulting from the combination of rotation and expansion, we show in the right panel of Figure \ref{fig7} trajectories calculated on a thin disc that has an axis projecting on the y axis and making an angle $i$=70\dego\ with the line of sight for different values of the ratio between $V_{exp}$ and $V_{rot}$.

\begin{figure}
  \includegraphics[width=0.46\textwidth,trim=0.cm 0.5cm 0 0.,clip]{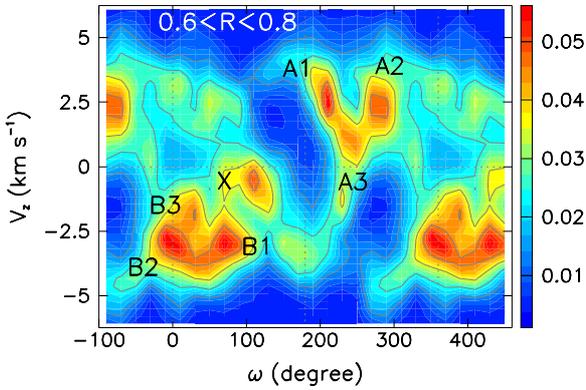}
  \caption{P-V map in the $V_z$ vs $\omega$ plane for projected distances from the star in the interval $0.6<R<0.8$ arcsec. The colour scale is in units of Jy beam$^{-1}$.}
  \label{fig8}
\end{figure}

\begin{figure*}
  \includegraphics[height=0.7\textheight,trim=0.0cm 0.cm 1.5 1.5,clip]{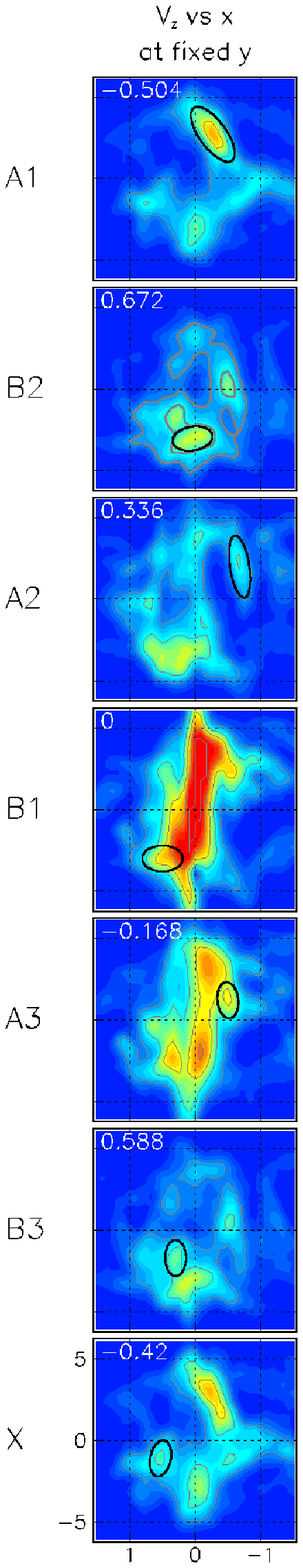}
  \includegraphics[height=0.7\textheight,trim=1.2cm 0.cm 1.5 1.5,clip]{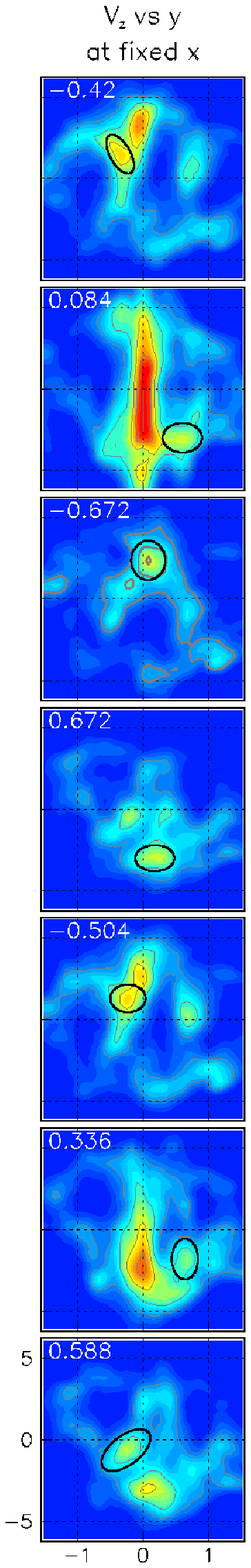}
  \includegraphics[height=0.7\textheight,trim=1.2cm 0.cm 1.5 1.5,clip]{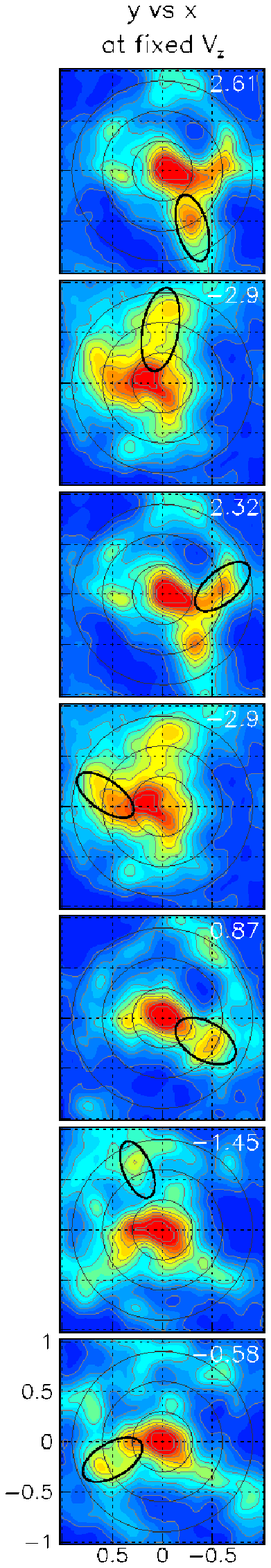}
  \includegraphics[height=0.7\textheight,trim=1.2cm 0.cm 1.5 1.5,clip]{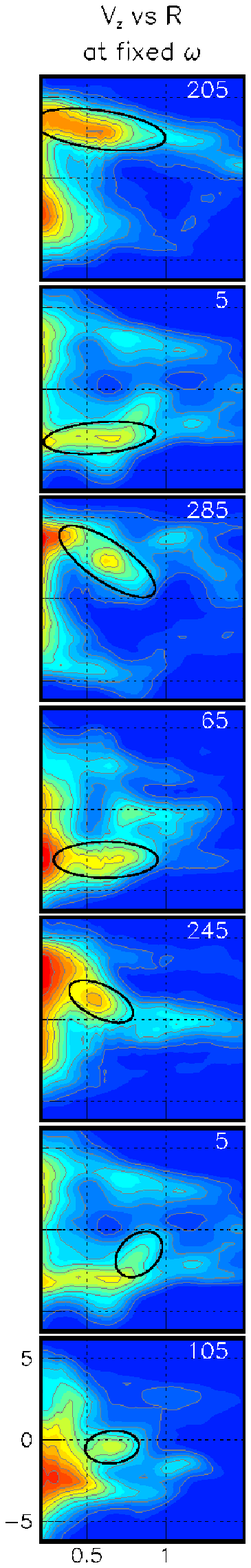}
  \caption{Projections on different planes of the data-cube environment of the outflows identified in Figure \ref{fig8}. The values of $y$, $x$, $V_z$ and $\omega$ are given in the inserts. Angles are in degrees, distances in arcsec and velocities in \kms.}
  \label{fig9}
\end{figure*}

In such a context, it is difficult to identify outflows reliably. Close to the star, the effect of rotation is too important and far from the star the outflows, more precisely the emission of the SO line, fade away rapidly. We therefore choose tentatively the interval of projected distances from the star $0.6<R<0.8$ arcsec as best compromise. Figure \ref{fig8} displays the associated P-V map in the $V_z$ vs $\omega$ plane. It shows (Table \ref{tab2}) four strong blobs of emission, labelled A1, A2, B1 and B2, and three fainter ones, labelled A3, B3 and X. The morpho-kinematics in their environment is illustrated in Figure \ref{fig9}. It shows approximate symmetry of the (A1,A2) and (B1,B2) pairs about a position angle of $\sim140$\dego$\pm$5\dego\ and a Doppler velocity of $\sim-0.3\pm0.2$ \kms, the separation between them being $\sim200$\dego\ and $\sim5.4$ \kms. The position angle of the approximate symmetry axis, $\sim140$\dego, differs clearly from the position angle of the rotation axis in the inner region, $\sim$20\dego, suggesting that outflows and inner rotation are unrelated. It matches instead the position angle of the blue-shifted high Doppler velocity component, which \citet{Decin2018} and \citet{Homan2018} call the ``blue-blob'', probably an accidental coincidence.   

\begin{table}
  \caption{Positions of the seven outflows identified in Figure \ref{fig8} (the numbers behind the $\pm$ signs give the approximate extensions of the outflows).}
  \label{tab2}
  \centering
  \begin{tabular}{ccc}
    \hline
    Name&$\omega$ (\dego)&$V_z$ (\kms)\\
    \hline
    A1&208$\pm$12&2.5$\pm$1.0\\
    \hline
    A2&280$\pm$12&2.2$\pm$0.8\\
    \hline
    A3&242$\pm$10&1.0$\pm$0.5\\ 
    \hline
    B1&73$\pm$18&$-$3.0$\pm$0.8\\
    \hline
    B2&5$\pm$25&$-$3.0$\pm$0.8\\
    \hline
    B3&25$\pm$7&$-$1.5$\pm$0.5\\
    \hline   
    X&110$\pm$10&$-$0.7$\pm$0.5\\
    \hline
    (A1+B1)/2&141$\pm$16&$-$0.2$\pm$0.9\\
    \hline
    (A2+B2)/2&143$\pm$18&$-$0.4$\pm$0.8\\
    \hline
    (A3+B3)/2&133$\pm$8&$-$0.2$\pm$0.5\\
    \hline
    {\it Mean}&139$\pm$14&$-$0.3$\pm$0.7\\
    \hline        
  \end{tabular}
\end{table}

A common property of these outflow candidates is to display an enhanced intensity at projected distances between 0.5 and 1 arcsec, suggesting an enhanced episode of mass loss a century or so ago (1 arcsec/century$\sim$3 \kms) as remarked earlier when commenting on Figure \ref{fig4}. This is further discussed below in Section \ref{sec4.3}.

In summary, evidence has been found for two radial outflows dominating the morpho-kinematics of the circumstellar envelope between $\sim$0.5 and 1 arcsec projected distance from the star; one is red-shifted and the other blue-shifted, with Doppler velocities of $\sim$2.4 and $-$3.0 \kms, approximately symmetric about a position angle of $\sim$140\dego; they are not quite back to back but $\sim$200\dego\ apart rather than 180\dego. Each of these outflows is dominated by two cores of nearly equal Doppler velocities, approximately 70\dego\ apart in position angle (A1, A2, B1 and B2). Fainter emissivity enhancements, labelled A3, B3 and X are detected as appendices of the main cores at lower Doppler velocities (probably closer to the plane of the sky).

\subsection{Central rotation}\label{sec4.2}

\begin{figure*}
  \includegraphics[height=0.30\textheight,trim=0.2cm .7cm 2. 1.,clip]{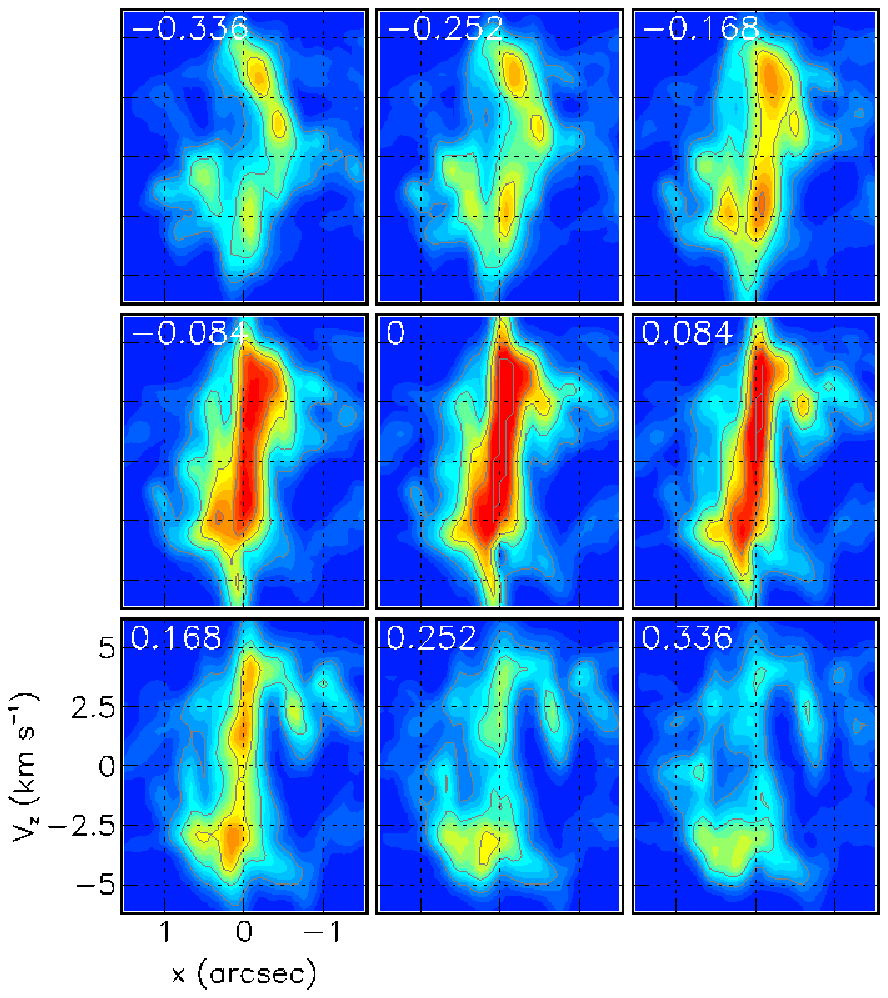}
  \includegraphics[height=0.30\textheight,trim=0.8cm .7cm 1.5 1.,clip]{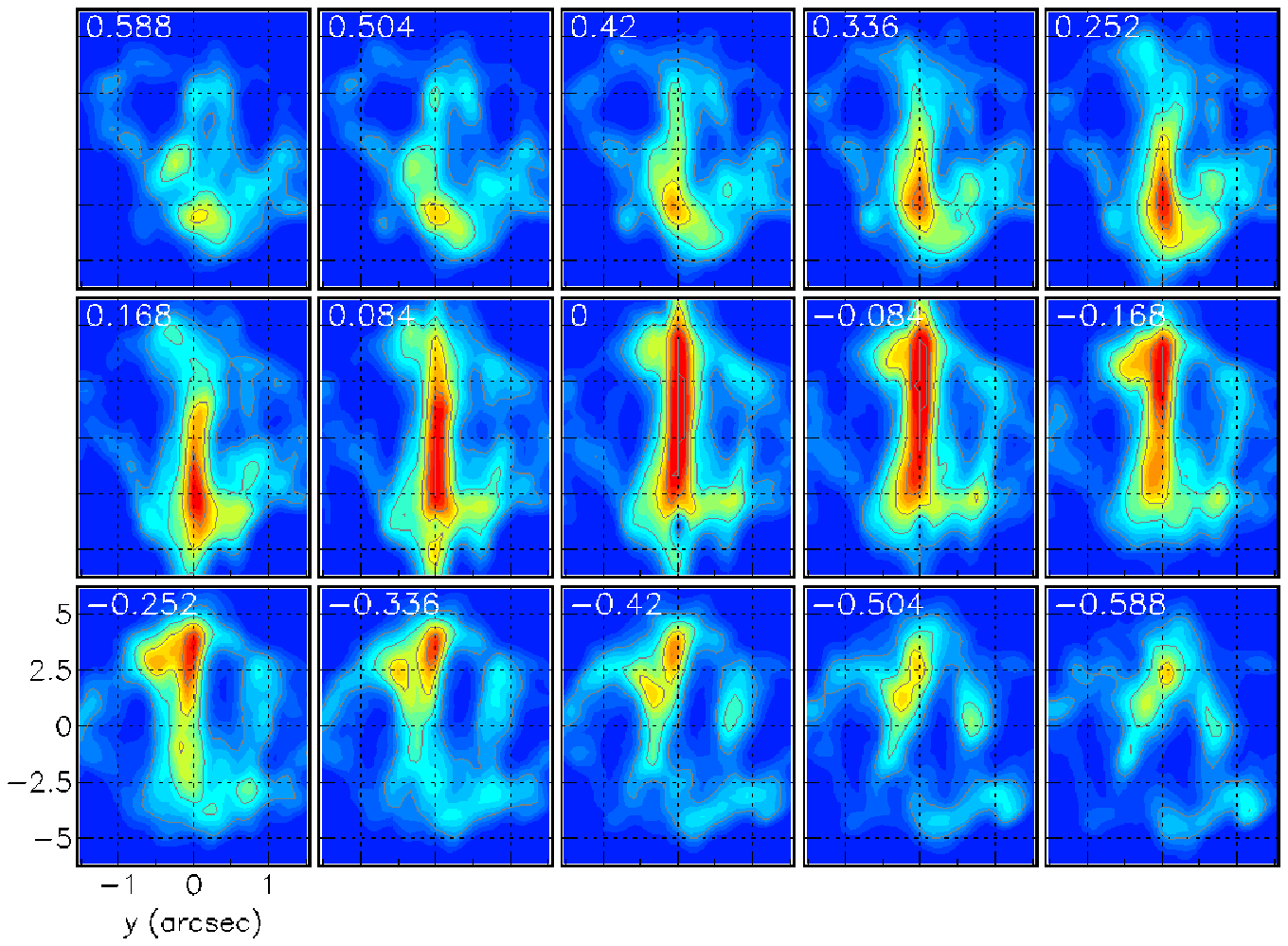}
   \includegraphics[height=0.25\textheight,trim=0.cm .5cm 1.5 2.,clip]{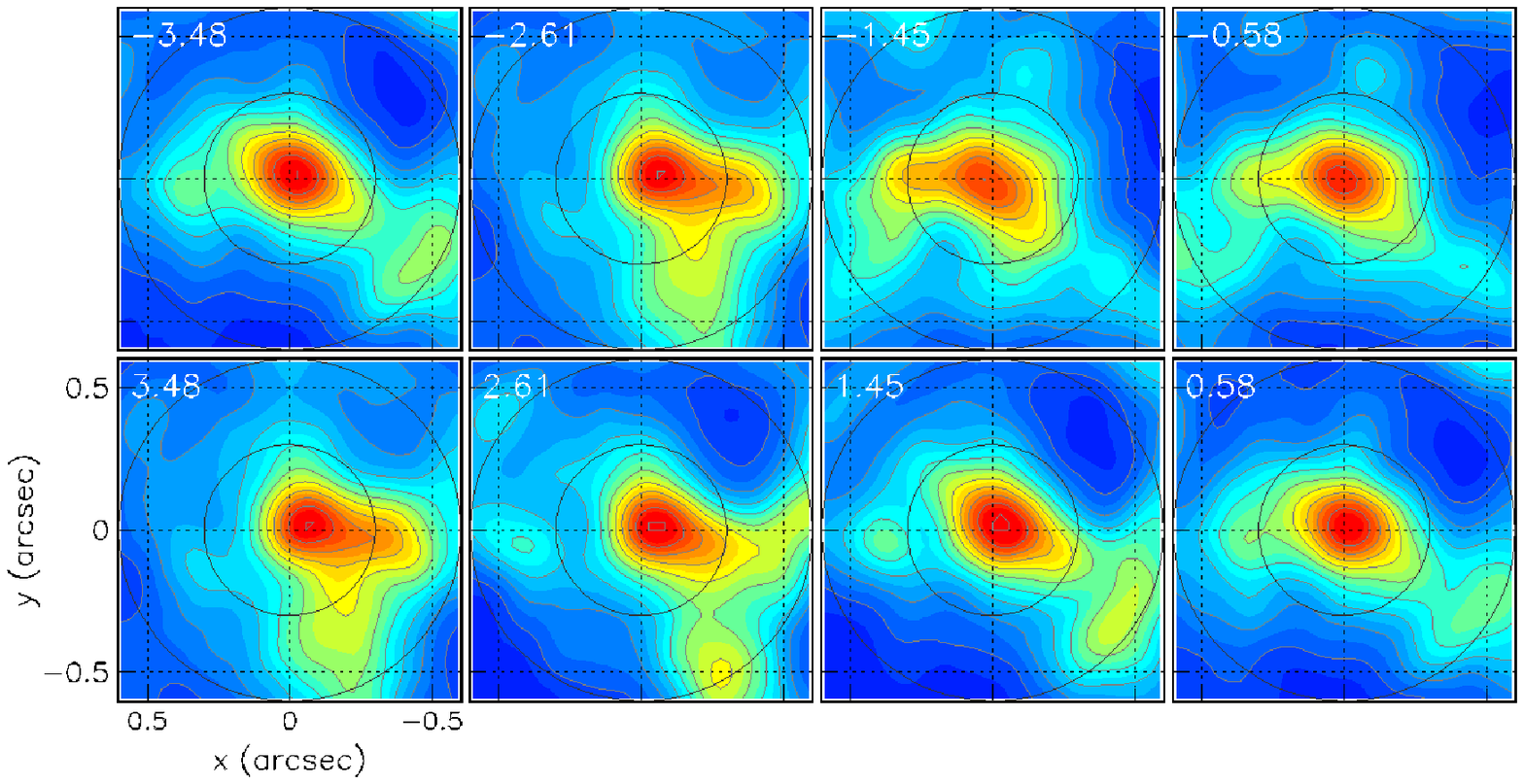}
  \caption{The data-cube near the line of sight. Upper panels: $V_z$ vs $x$ in steps of $y$ (three leftmost columns) and $V_z$ vs $y$ in steps of $x$ (five rightmost columns). Lower panels: $x$ vs $y$ at selected Doppler velocities near $\pm$0.5, $\pm$1.5, $\pm$2.5 and $\pm$3.5 \kms.}
  \label{fig10}
\end{figure*}

Having found evidence for the presence of radial outflows in the preceding section, we dedicate the present section to an inspection of the data-cube in the central region, which \citet{Homan2018} have shown to host a rotating nearly edge-on disc. The detailed morphology and kinematics of the emission detected near the line of sight is best studied on the ($V_z$ vs $x$) and ($V_z$ vs $y$) P-V maps displayed in Figure \ref{fig10}. Emission is confined to the three to five central panels of the former, covering $\sim\pm0.2$ arcsec in $y$. This is consistent with the model of a disc rotating about an axis projecting close to the $y$ axis and making a small angle with the plane of the sky, as advocated by \citet{Homan2018}. Indeed, the ($V_z$ vs $y$) P-V maps cover a range of $\sim\pm0.5$ arcsec in $x$, providing a measure of the outer disc radius, $\sim$0.5 arcsec, while the ratio 0.2/0.5$\sim$0.4 provides a measure of the inclination angle, $\sim$20\dego. These estimates are in good agreement with the results of \citet{Homan2018}, 25 au ($\sim$0.4 arcsec) and 20\dego\ respectively.

When scanning in $x$ from east, one first meets $V_z$ values of $-$3.5 to $-$2.5 \kms, reached at $x\sim0.5$ arcsec and $y\sim0$, measuring a tangential rotation velocity of $\sim$3 \kms\ at the disc edge and implying that the disc rotates clockwise when seen from south; at this point, a possible radial expansion, being perpendicular to the line of sight, would have no effect. One expects a symmetric situation when scanning from west, namely to first meet  $V_z$ values centred on $\sim$+3 \kms\ at $x\sim-0.5$ arcsec and $y\sim0$. This is indeed what is observed but the picture is now confused by the presence of a blob of emission  that is easily identified as coming from the southern core of the red-shifted enhancement, A1, studied in the preceding section. However, when reaching $x\sim0$, where rotation velocities are perpendicular to the line of sight, one would expect to find an enhancement of emission confined near the origin, which is not at all the case; but this is the region where the black body radiation of the star is important and where the spatial resolution prevents a reliable study of the line emission.

When scanning in $y$ from south or from north, the beam size (0.15 arcsec FWHM) and the disc size, $\sim\pm$0.2 arcsec, prevent as fine a sampling as was possible when scanning in $x$. Yet, one would expect to first meet zero values of both $x$ and $V_z$, which is not the case: on the contrary, emission is enhanced at both $V_z=\pm3$ \kms, whether $y$ approaches zero from above or from below, requiring important radial expansion. But this, again, is close to the central region where a reliable quantitative evaluation is not possible. Within $\pm\sim$8 au from the star (the three central panels in the $V_z$ vs $x$ maps of Figure \ref{fig10}), $V_z$ is negative east and positive west, in agreement with the rotating disc model.

In summary, inspection of the central region of the data cube confirms the presence of a rotating disc as found by \citet{Homan2018} but gives also evidence for an important contribution of radial expansion, probably associated with the nascent outflows studied in the preceding section.

\subsection{A toroidal cavity}\label{sec4.3}

\begin{figure*}
  \includegraphics[width=0.3\textwidth,trim=0.5cm 1.cm 0 0.5,clip]{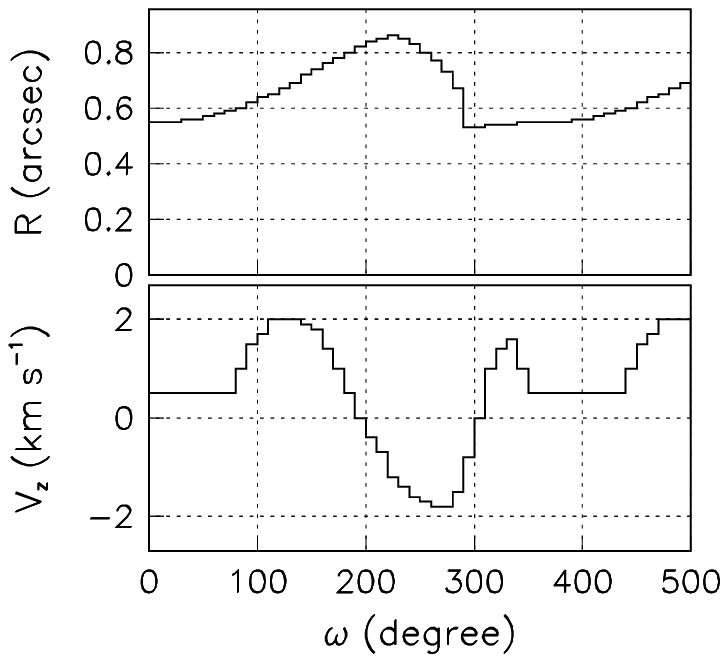}
  \includegraphics[width=0.3\textwidth,trim=0.5cm 1.cm 0 0.5,clip]{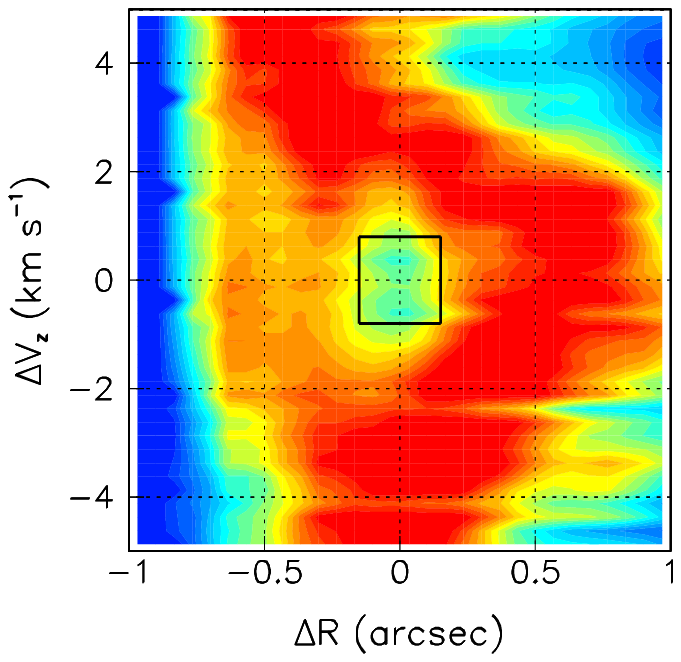}
  \includegraphics[width=0.3\textwidth,trim=0.5cm 1.cm 0 0.5,clip]{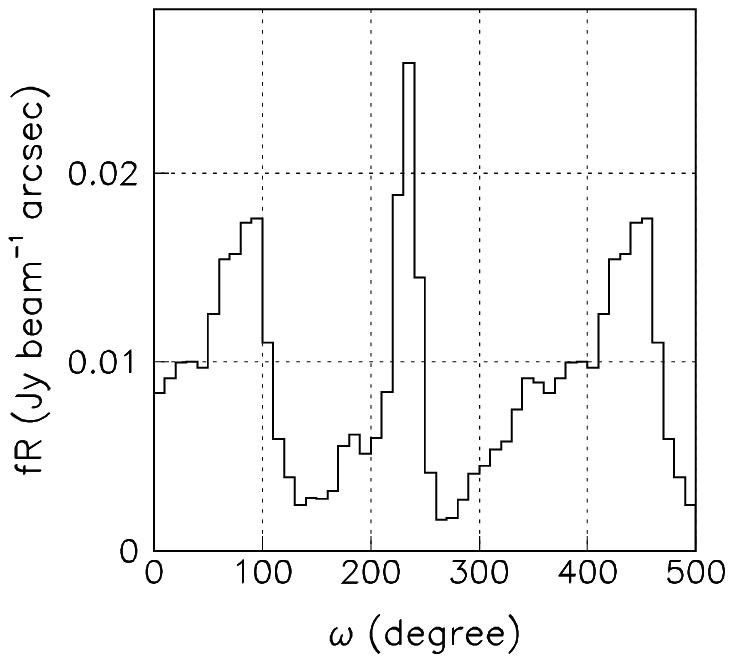} 
  \caption{Left: Dependence on $\omega$ (degree) of $R_{cent}(\omega)$ (arcsec, upper panel) and $V_{zcent}(\omega)$ (\kms, lower panel) as estimated from a detailed inspection of the data-cube in the K1 region. Middle: Distribution of $fR$ as a function of $R- R_{cent}(\omega)$ (abscissa) and $V_z-V_{zcent}(\omega)$ (ordinate), averaged over $\omega$. Right: dependence on $\omega$ of $fR$ averaged over the cavity, as defined by the square ($\pm$0.15 arcsec, $\pm$0.8 \kms) shown in the middle panel.}
  \label{fig11}
\end{figure*}

As already mentioned in Section \ref{sec3} the data-cube seems to host two ring-shaped depletions labelled K1 and K2 in Figure \ref{fig4}. Indeed, a detailed inspection of the K1 region shows the presence of a ring cavity that runs around all position angles $\omega$, centred at values of ($R$,$V_z$) called ($R_{cent}(\omega)$,$V_{zcent}(\omega)$) that vary with $\omega$ as displayed in the left panels of Figure \ref{fig11}. They correspond approximately to a torus inclined with respect to the plane of the sky, reaching extreme Doppler velocities of nearly $\pm2$ \kms\ at $\omega\sim$110\dego\ on the red-shifted side and $\omega\sim290$\dego\ on the blue-shifted side. The distribution of the product of the brightness $f$ by the projected distance $R$ as a function of $R-R_{cent}(\omega)$ and $V_z-V_{zcent}(\omega)$, averaged over $\omega$, is shown in the middle panel of the figure, giving clear evidence for the presence of the cavity. The dependence on $\omega$ of product $Rf$, averaged over the cavity, is shown on the right panel of the figure. It shows emission at values of $\omega$ corresponding with the crossing of the cavity by the low Doppler velocity part of the blue-shifted and red-shifted outflows identified in the preceding section.

\begin{figure}
  \includegraphics[width=0.48\textwidth,trim=0.cm 1.cm 1. 0.5,clip]{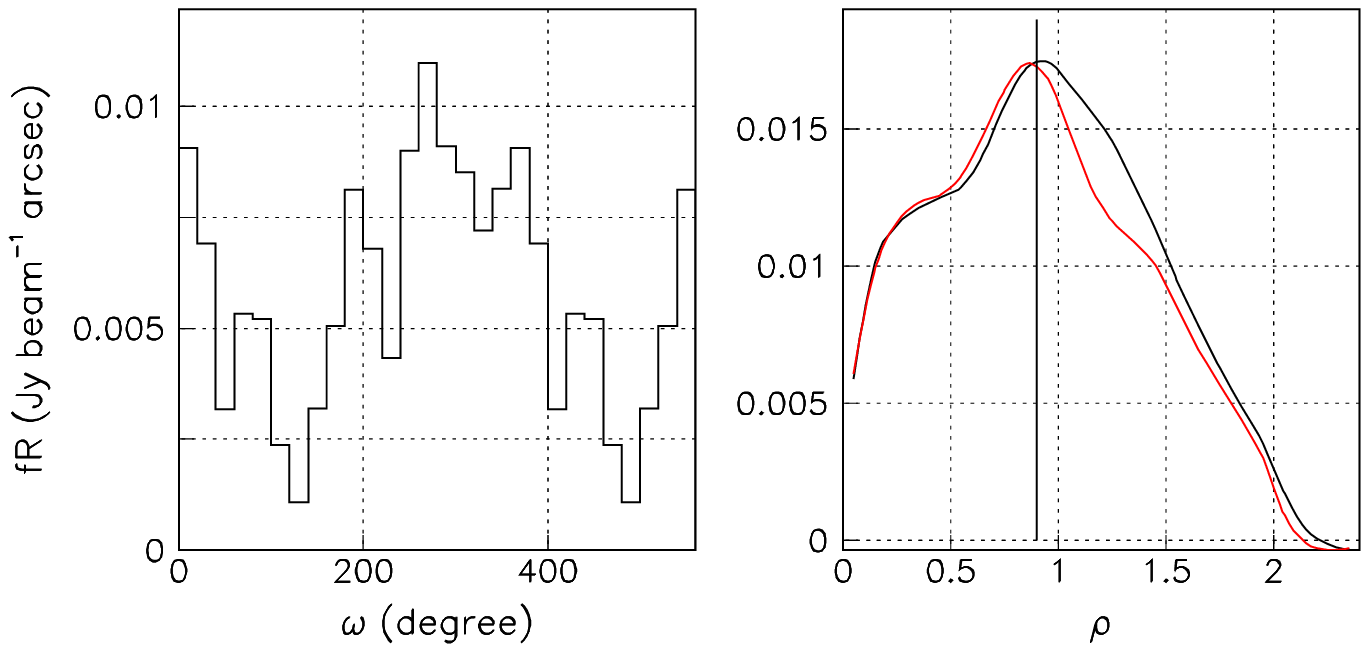}
  \caption{Left: Dependence on $\omega$ of the product $Rf$ for $R>1.2$ arcsec. Right: distribution of product $Rf$ as a function of $\rho$ for $V_0=4$ (red) and 5 (black) \kms.}
  \label{fig12}
\end{figure}

In principle, such a cavity could have been carved in the mass-losing wind by a companion rotating about the central star. However, in view of what has been learned in the preceding sections, it is more natural to associate it with the period of more quiet mass loss that followed the episode of enhanced emission that has been suggested to have occurred a century or so ago. The presence of a second ring cavity, labelled K2 in Figure \ref{fig4}, at projected distances in excess of $\sim1.2$ arcsec might then suggest that another period of enhanced emission occurred another century before the more recent one. However, its location in a region where the SO abundance declines rapidly and where the signal-to-noise ratio becomes small prevents a reliable study and its existence cannot be assessed with confidence. We simply show, in the left panel of Figure \ref{fig12}, the dependence on $\omega$ of the mean value of product $Rf$ for $R>1.2$ arcsec; it displays a clear minimum in the south-eastern hemisphere near $\omega\sim140$\dego.

As further evidence for the time dependence of the mass loss, Figure \ref{fig12} displays the product $Rf$ averaged over the data cube as a function of the parameter $\rho=\sqrt{R^2+V_z^2/V_0^2}$ measuring the distance in the data cube to an ellipsoid reaching $\pm V_0$ on the $V_z$ axis and cutting the plane of the sky as a circle of radius 1 arcsec. As remarked earlier in Section \ref{sec3} when commenting on the right panel of Figure \ref{fig4}, one expects a value $V_0$ between 4 and 5 \kms\ to give evidence for a maximum of emission near $\rho\sim1$. Figure \ref{fig12} (right panel) displays the dependence of product $Rf$ on $\rho$ for both values of $V_0$, 4 and 5 \kms, showing a clear excess near $\rho\sim0.9$. 

\subsection{Comparison with other molecular lines} \label{sec4.4}

The circumstellar envelope of R Dor has been probed at ALMA using several other molecular line emissions, as described by \citet{Decin2018}. In \citet{Hoai2019} we have made use of some of these observations to comment on the high Doppler velocity components of the nascent wind. In the present section we compare them with the results obtained in the preceding sections from the analysis of SO($J_K=6_5-5_4$) emission. In spite of much shorter times on source, of higher noise level and of an antenna configuration implying a short reach in $R$ they provide useful information to complement and validate the results obtained on the emission of the SO line.

\begin{figure*}
  \includegraphics[width=0.75\textwidth,trim=0.cm 0.5cm 0.5 0.,clip]{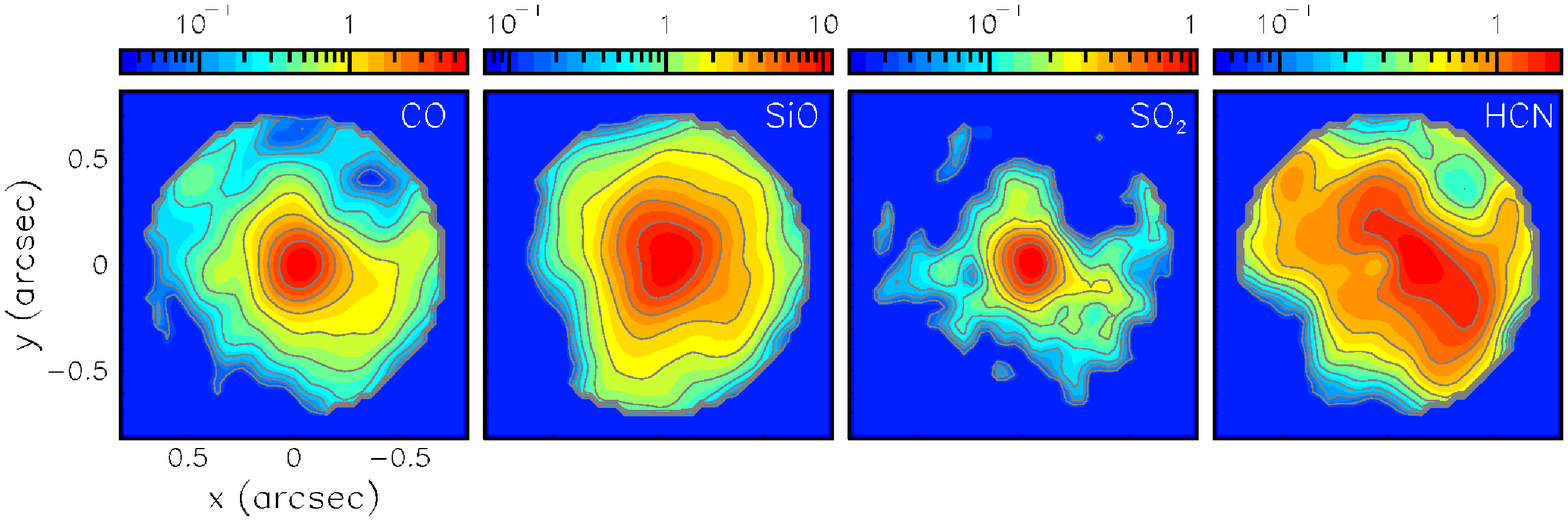}
  \caption{Intensity maps of the red-shifted hemispheres for each of the CO, SiO, SO$_2$ and HCN lines. The colour scales are in units of Jy beam$^{-1}$ \kms.}
  \label{fig13}
\end{figure*}

The data are retrieved from ALMA archives and have been reduced and continuum subtracted by the ALMA staff. They are from project 2013.1.00166.S observed in summer 2015 in band 7 with an average of 39 antennas (Table \ref{tab3}). Their interpretation in the blue-shifted hemisphere is complicated by the presence of strong absorption of the black-body emission of the stellar disc (for a detailed discussion see Appendix B of \citet{Wong2016} who discuss a similar problem in the case of Mira Ceti): intensity maps limited to the red-shifted hemisphere are shown in Figure \ref{fig13} for each of the four lines, CO, SiO, SO$_2$ and HCN. All but the SiO map display major departure from isotropy with an elongation in the north-eastern to south-western direction, consistent with the approximate symmetry about a position angle of $\sim$140\dego\ observed for SO emission. The maximal projected distance from the star within which reliable information can be obtained does not exceed 40 au. However, while a detailed and critical study of these observations is beyond the scope of the present article, important information contained at projected distances from the star between $\sim$15 and $\sim$40 au deserves being briefly commented upon.

\begin{table*}
  \caption{Line emissions considered in Section \ref{sec4.4}. Beam FWHM is in mas$^2$ and noise in mJy beam$^{-1}$.}
  \label{tab3}
  \begin{tabular}{ccccc}
    \hline
    Line&CO(3-2)&SiO(8-7)&SO$_2$(13$_{4,10}$-13$_{3,11}$)&HCN(4-3)\\
    \hline
    Beam&180$\times$140&180$\times$130&160$\times$130&157$\times$145\\
    Noise&5.5&4.8&6.3&8.5\\
    \hline
  \end{tabular}
\end{table*}

\begin{figure*}
  \includegraphics[width=0.67\textwidth,trim=0.cm 0.5cm 0.5 0.5,clip]{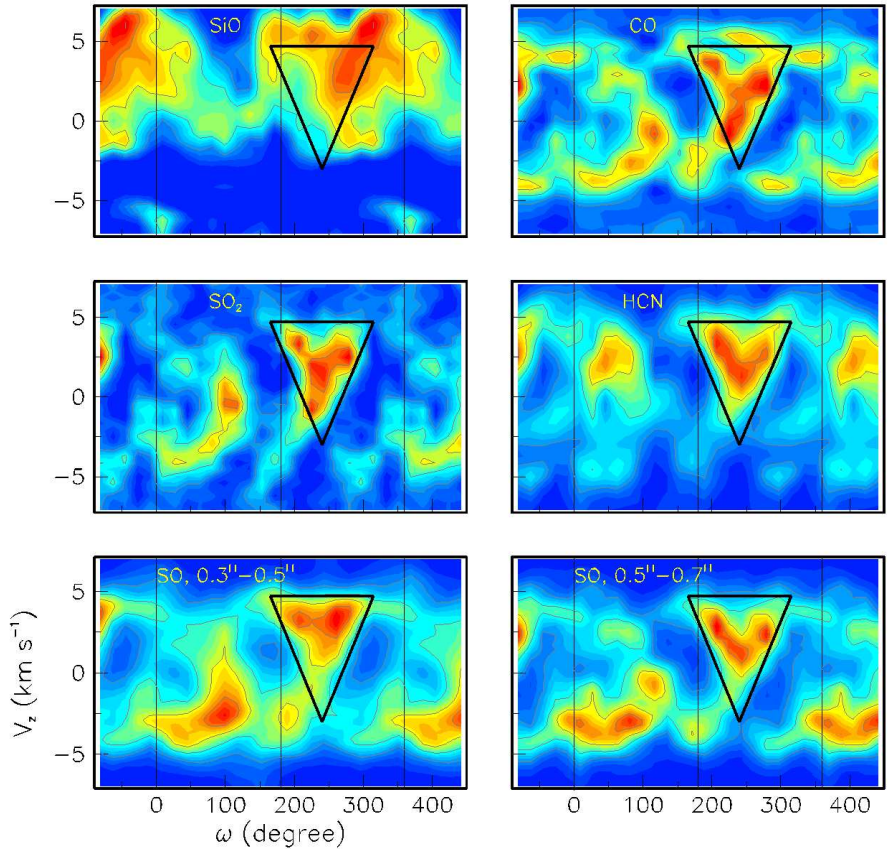}
  \caption{P-V maps in the $V_z$ vs $\omega$ plane for projected distances from the star in the interval $0.3<R<0.7$ arcsec. Lines are identified in the inserts. For the SO line, the interval of $R$ is split in two equal parts. A same triangle embedding the A outflows is shown on each map to guide the eye. }
  \label{fig14}
\end{figure*}

Figure \ref{fig14} displays P-V maps in the $V_z$ vs $\omega$ plane for projected distances from the star in the interval $0.3<R<0.7$ arcsec. The P-V maps of the CO, SO$_2$ and HCN lines display patterns similar to that observed  for the SO line. In particular outflows A1, A2 and A3 that were identified in Figure \ref{fig8} are observed to dominate the western hemisphere at approximately the same values of $\omega$ and $V_z$. However, in the eastern hemisphere, the observed emission tends to be closer from the X outflow than from the B outflows, the latter being visible on the CO map only. Two main parameters govern the morphology of the observed patterns: the distance from the star over which the abundance of the emitting molecule is significant and the projected distance on the sky plane over which the acceptance of the interferometer is significant. The former, together with parameters on which the emissivity depends, such as the gas temperature, exerts its influence over the whole line of sight, independently from the latter: it decides over which distance the envelope is being probed on average, possibly outside the range of the interferometer acceptance. Comparing the SO$_2$ pattern with  the SO pattern observed at shorter values of $R$, as displayed in the right panels of Figure \ref{fig5} and the lower left panel of Figure \ref{fig14}, we see that they are similar, indicating that SO$_2$ is probably probing shorter distances from the star than CO and SO are. In the case of SiO, where absorption affects the whole blue-shifted hemisphere, one observes a radically different distribution when compared with the other lines, filling much of the cavities. Moreover, it reaches higher Doppler velocities than the other lines, suggesting that it may probe preferentially distances from the star where the slow wind experiences stronger acceleration.

\begin{figure*}
  \includegraphics[width=0.9\textwidth]{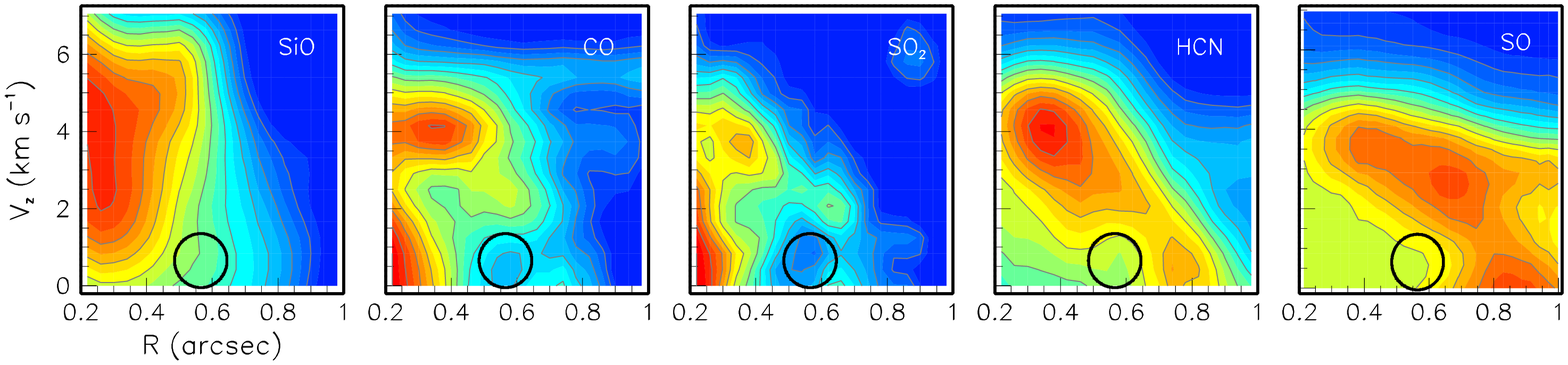}
  \caption{P-V maps of the product of the brightness $f$ by the projected distance $R>0.2$ arcsec in the $V_z>0$ vs $R$ plane, averaged over position angle $\omega$. Lines are identified in the inserts. A same circle locating cavity K1 is shown on each panel to guide the eye.}
  \label{fig15}
\end{figure*}

Figure \ref{fig15} displays P-V maps of the product of the brightness $f$ by the projected distance $R$ in the $V_z>0$ vs $R>0.2$ arcsec  plane, averaged over position angle $\omega$. Cavity K1 is clearly visible on most maps but is filled up on the SiO map. 

\begin{figure*}
  \includegraphics[width=0.7\textwidth,trim=0cm 1.5cm 0 0,clip]{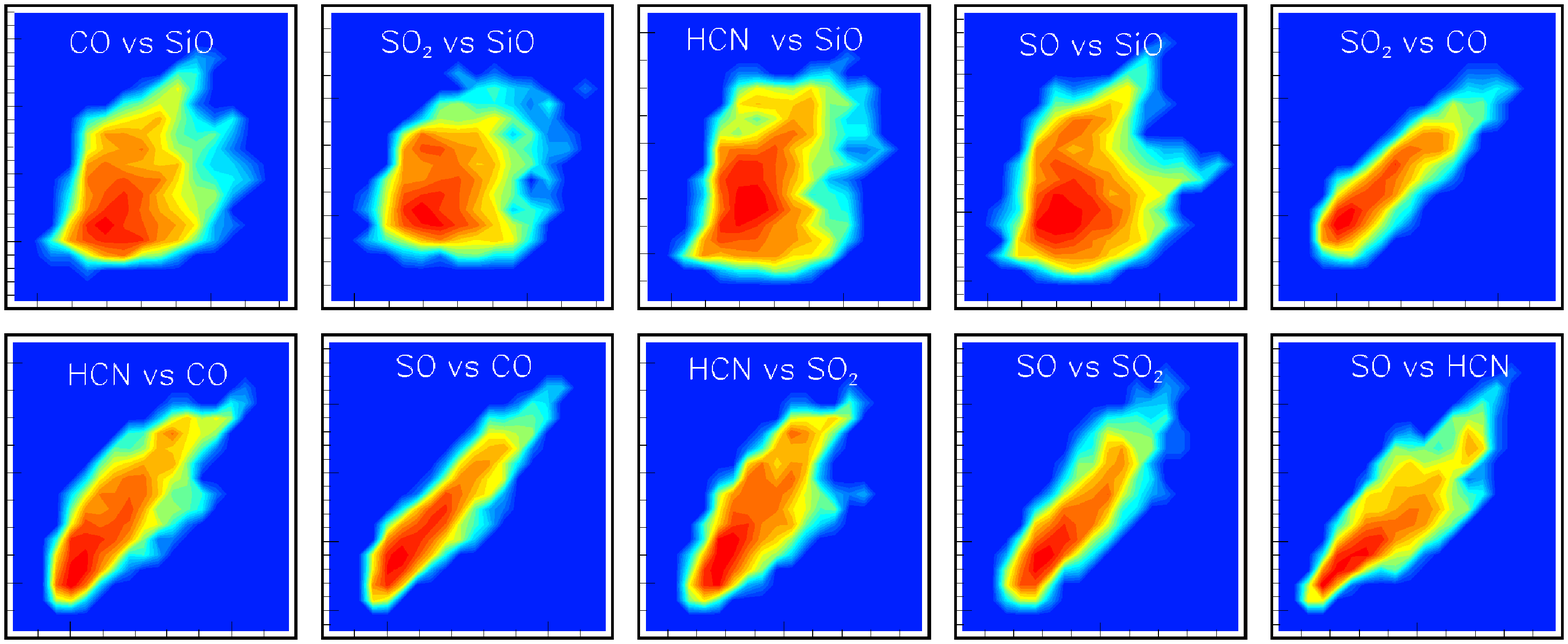}
  \includegraphics[width=0.7\textwidth,trim=0cm 1.5cm 0 0,clip]{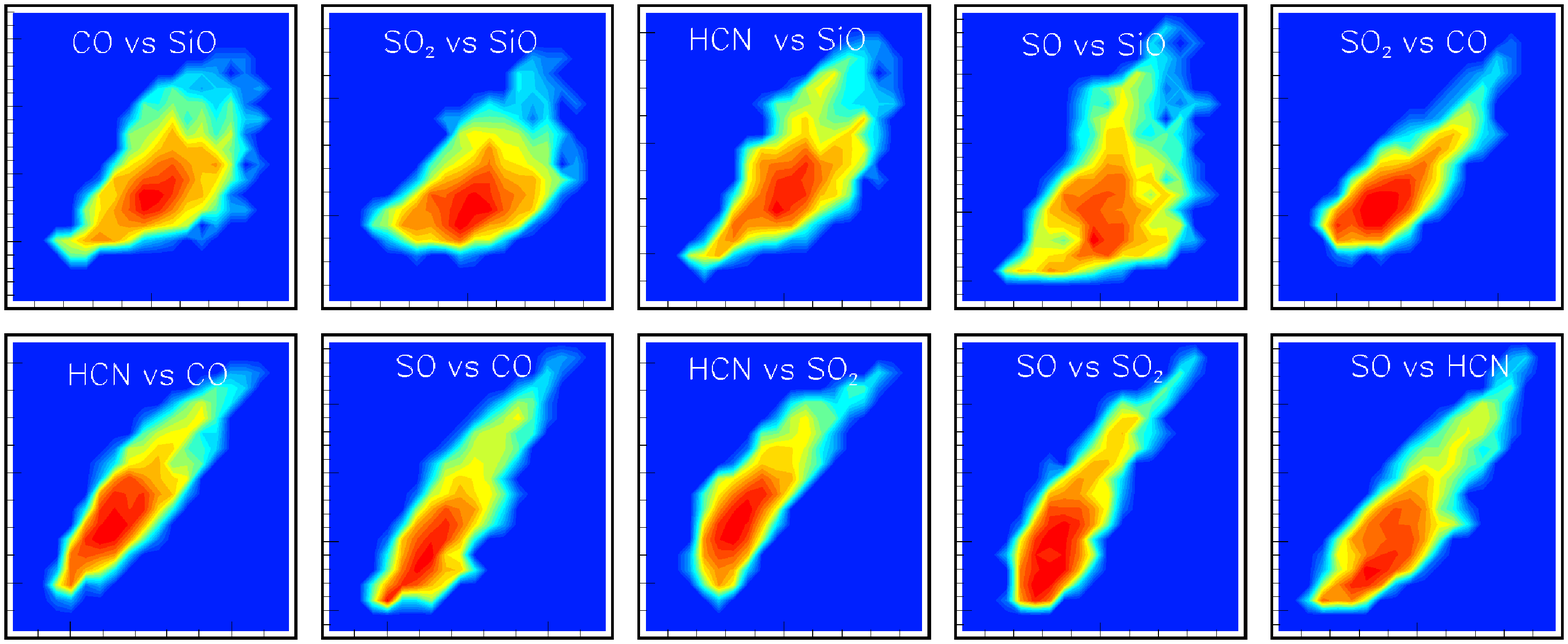}
  \caption{Correlation between measured brightness in the interval $0.3<R<0.5$ arcsec and $0<V_z<3$ \kms\ (upper rows) or $3<V_z<6$ \kms\ (lower rows). Lines are identified in the inserts. The scales of abscissa and ordinate are adjusted in each case to cover the whole distribution.}
  \label{fig16}
\end{figure*}

As a further comparison between the SO results and the emission of the other lines, we display in Figure \ref{fig16} the correlation between the measured brightness in the region $0.3<R<0.5$ arcsec and $0<V_z<6$ \kms\ where reliable measurements are available for all lines. A clear positive correlation is found between the SO$_2$, HCN, CO and SO data while their correlation with the SiO data is much weaker. By splitting the Doppler velocity interval in two equal parts, one sees that the disagreement between the SiO data and the other lines is strongly enhanced at lower Doppler velocities.     

We have checked that this disagreement is similarly present in another SiO data set from the same project that we have reduced using ALMA pipeline standard script, GILDAS and natural weighting and have not continuum subtracted. The resulting beam size is 220$\times$170 mas$^2$ and the noise is 5.2 mJy beam$^{-1}$. The disagreement is therefore not an effect of $uv$ coverage or data reduction. \citet{Decin2018} have commented on the nucleation process of dust grains in the radial range of relevance (their Section 5.3.2) and a detailed modelling, probably out of reach of current understanding, would be required to explain the difference between the observed morpho-kinematics of SiO and that of other molecular species. 

\section{Summary}\label{sec5}

In spite of the complexity of the picture that emerges from the present analysis of the data-cube of SO($J_K=6_5-5_4$) emission, a number of new reliable results have been obtained with reasonable confidence. The good match between the distances covered by the SO abundance and by the acceptance obtained from an appropriate \textit{uv} coverage and a long enough time on source have made it possible.

Within projected distances $R<\sim30$ au, the morpho-kinematics is consistent with the presence of a disc rotating clockwise when seen from south about an axis projecting $\sim20$\dego\ east of north and making an angle of $\sim20$\dego\ with the plane of the sky; the present SO observations fully confirm the results of \citet{Homan2018} who were first to identify the disc as having a radius of $\sim$25 au, a tangential rotation velocity of $\sim$3 \kms\ at the outer edge and a thickness of a few au.

Projected distances between $\sim$20 and $\sim$100 au host a slow radial wind with Doppler velocities reaching up to $\sim$6 \kms; it is characterized by strong inhomogeneity, both in direction and radially, as had been noted earlier by \citet{DeBeck2018}. The former takes the form of separate cores, together covering very large solid angles, and the latter suggests the occurrence of an episode of enhanced mass loss at the scale of a century ago. At short distances from the star the combined effect of expansion and rotation makes a reliable analysis of the morpho-kinematics difficult. Two radial outflows dominate the morpho-kinematics of the circumstellar envelope between $\sim0.5$ and 1 arcsec projected distance from the star ($\sim30$ and 60 au); one is red-shifted and the other blue-shifted, with Doppler velocities of $\sim2.4$ and $-3.0$ \kms, approximately symmetric about a position angle of $\sim140$\dego; they are not quite back to back but $\sim200$\dego\ rather than 180\dego\ apart. Each of these outflows is dominated by two cores of nearly equal Doppler velocities, approximately 70\dego\ apart in position angle and fainter emissivity enhancements are detected as appendices of the main cores at lower Doppler velocities.

\begin{figure}
  \includegraphics[width=0.4\textwidth]{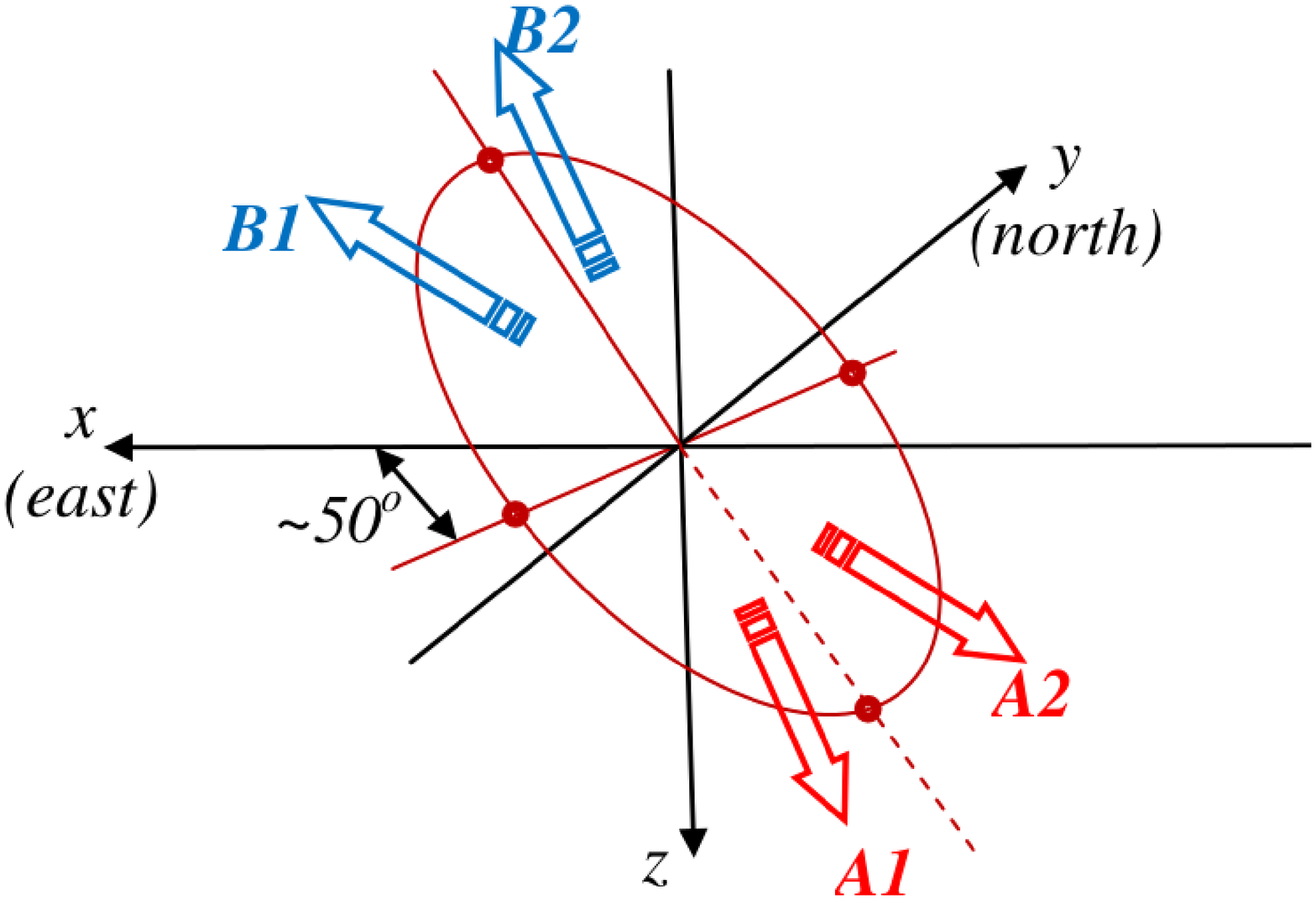}
  \caption{Qualitative picture summarizing the results obtained in the present work.}
  \label{fig17}
\end{figure}

Figure \ref{fig17} gives a qualitative picture that summarizes the present results. Assuming that the four main cores of emission have a common radial velocity of $\sim9$ \kms\ and are therefore nearly coplanar, the inclination of their plane with respect to the plane of the sky is $\sim\sin^{-1}(3/9)\sim20$\dego. For a common radial velocity of 6 \kms, the inclination becomes 30\dego.

Emission from other molecular lines, detected in less favourable observational conditions than the SO line, normally confirms the presence of the radial outflows identified here and gives evidence for different molecular species probing preferentially different distances from the star. However, emission of the SiO line displays a significantly different pattern, suggesting that SiO gas populates and probes significantly different regions of the circumstellar envelope than do the other lines.  

High spatial resolution ALMA observations of oxygen-rich AGB stars are often revealing a much more complex morpho-kinematics than had been originally anticipated. It has been the case with Mira for now many years, recent examples are QX Pup \citep{SanchezContreras2018,Olofsson2019}, R Aqr \citep{Ramstedt2018} and R Leo \citep{Fonfria2019}, raising new questions that are still awaiting an answer. In the present case of R Dor, several such questions are present: the motor of the inner rotation providing the required angular momentum is unknown; \citet{Vlemmings2018} and \citet{Homan2018} argue for the presence of a close by companion but we have no evidence for its existence and a magnetic field origin is another possibility \citep{Matt2000}; a detailed study of the strong absorption observed in the blue-shifted hemisphere, in particular for the SiO emission line, possibly including a contribution from in-falling gas, needs still to be done; the dynamics underlying the emission of the radial outflows revealed in the present work is not understood; the nature of the high Doppler velocity components, apparently unrelated to other observations, is unknown. We are far from being able to tell a consistent story describing the nascent wind in the circumstellar envelope of this AGB star and obeying the constraints imposed by hydrodynamics.

\section*{ACKNOWLEDGEMENTS}

We thank Prof. Leen Decin and Dr. Ward Homan for having shared with us their understanding of the inner wind of R Dor. This paper uses ALMA data 2017.1.00824.S and 2013.1.00166.S. ALMA is a partnership of ESO (representing its member states), NSF (USA), NINS (Japan), NRC (Canada), NSC/ASIAA (Taiwan), and KASI (South Korea), in cooperation with Chile. The Joint ALMA Observatory is operated by ESO, AUI/NRAO and NAOJ. The data are retrieved from the JVO/NAOJ portal. We are deeply indebted to the ALMA partnership, whose open access policy means invaluable support and encouragement for Vietnamese astrophysics. Financial support from the World Laboratory and VNSC is gratefully acknowledged. This research is funded by Vietnam National Foundation for Science and Technology Development (NAFOSTED) under grant number 103.99-2018.325.





\begin{thebibliography}{99}

\bibitem[\protect \citeauthoryear {Bedding et al.}{1998}]{Bedding1998}
  Bedding T. R., Zijlstra A.A., Jones A., \& Foster G., 1998, MNRAS, 301, 1073
\bibitem[\protect \citeauthoryear {Danilovich et al.}{2016}]{Danilovich2016} 
  Danilovich T., De Beck E., Black J.H., et al., 2016, A\&A 588, A119
\bibitem[\protect \citeauthoryear {De Beck \& Olofsson}{2018}]
{DeBeck2018}
  De Beck E. \& Olofsson H., 2018, A\&A, 615, A8
\bibitem[\protect \citeauthoryear {Decin et al.}{2018}]{Decin2018}
  Decin L., Richards A.M.S., Danilovich T., et al., 2018, A\&A, 615, A28
\bibitem[\protect \citeauthoryear {Diep et al.}{2016}]{Diep2016}
Diep, P.N., Phuong, N.T., Hoai, D.T. et al., 2016, MNRAS, 461, 4276
\bibitem[\protect \citeauthoryear {Dumm \& Schild}{1998}]{Dumm1998}
  Dumm T., \& Schild H., 1998, New Astr., 3, 137
\bibitem[\protect \citeauthoryear {Fonfria et al.}{2019}]{Fonfria2019}
  Fonfria, J.P., Santander-Garcia, M., Cernicharo, J. et al., 2019, A\&A, 622, L14.
\bibitem[\protect \citeauthoryear {Heras \& Hony}{2005}]{Heras2005}
  Heras M. \& Hony S., 2005, A\&A, 439, 171
\bibitem[\protect \citeauthoryear {Hoai et al.}{2019}]{Hoai2019}
  Hoai D.T., Nhung P.T., Tuan-Anh P., et al., 2019, submitted to MNRAS
\bibitem[\protect \citeauthoryear {Homan et al.}{2018}]{Homan2018}
  Homan W., Danilovich T., Decin L., et al., 2018, A\&A, 614, A113
\bibitem[\protect \citeauthoryear {Khouri et al.}{2016}]{Khouri2016}
  Khouri T., Maercker M., Waters L.B.F.M., et al., 2016, A\&A, 591, A70
\bibitem[\protect \citeauthoryear {Knapp et al.}{2003}]{Knapp2003}
  Knapp G.R., Pourbaix D., Platais, I., and Jorissen, A., 2003, A\&A, 403, 993
\bibitem[\protect \citeauthoryear {Lebzelter \& Hron}{1999}]{Lebzelter1999}
  Lebzelter, T., and Hron, J., 1999, A\&A, 351, 533
\bibitem[\protect \citeauthoryear {Matt et al.}{2000}]{Matt2000}
  Matt S., Balick, B., Winglee, R. \& Goodson, A., 2000, ApJ, 545, 965
\bibitem[\protect \citeauthoryear {Nhung et al.}{2018}]{Nhung2018}
  Nhung P.T., Hoai D.T., Tuan-Anh P. et al., 2018, MNRAS, 480, 3324
\bibitem[\protect \citeauthoryear {Olofsson et al.}{2019}]{Olofsson2019}
  Olofsson, H., Khouri, T., Maercker, M. et al., 2019, A\&A, 623A, 1530
\bibitem[\protect \citeauthoryear {Ramstedt \& Olofsson}{2014}]{Ramstedt2014}
  Ramstedt, S., \& Olofsson H., 2014, A\&A 566, A145
\bibitem[\protect \citeauthoryear {Ramstedt et al.}{2018}]{Ramstedt2018}
Ramstedt, S., Mohamed, S., Olander, T. et al., 2018, A\&A, 616A, 61  
\bibitem[\protect \citeauthoryear {S\'{a}nchez Contreras et al.}{2018}]{SanchezContreras2018}
  S\'{a}nchez Contreras, C., Alcolea, J., Bujarrabal, V. et al., 2018, A\&A, 618, A164
\bibitem[\protect \citeauthoryear {TuanAnh et al.}{2015}]{TuanAnh2015}  
  Tuan Anh, P., Diep, P.N., Hoai, D.T. et al., 2015, RAA, 15, 2213
\bibitem[\protect \citeauthoryear {Tuan-Anh et al.}{2019}]{TuanAnh2019}
  Tuan-Anh P., Hoai D.T., Nhung P.T., et al., 2019, MNRAS, 487, 622
\bibitem[\protect \citeauthoryear {Van de Sande et al.}{2018}]{VandeSande2018}
  Van de Sande M., Decin L., Lombaert R., et al., 2018, A\&A 609, A63
\bibitem[\protect \citeauthoryear {van Leeuwen}{2007}]{VanLeeuwen2007}
  van Leeuwen F., 2007, A\&A, 474, 653
\bibitem[\protect \citeauthoryear {Vlemmings et al.}{2018}]{Vlemmings2018}
  Vlemmings W.H.T., Khouri T., De Beck E., et al., 2018, A\&A, 613, L4
\bibitem[\protect \citeauthoryear {Wong et al.}{2016}]{Wong2016}
  Wong K.T., Kaminski T., Menten K.M. \& Wyrowski F., 2016, A\&A, 590, A127

\end{thebibliography}





\bsp	
\label{lastpage}
\end{document}